\documentclass[prd,aps,a4paper,eqsecnum,floatfix,nofootinbib,twocolumn]{revtex4}  %

\usepackage{graphicx} 
\usepackage{amsmath,amsfonts,amssymb}
\usepackage{tikz}
\usepackage{appendix}
\usepackage[caption=false,justification=justified]{subfig}

\newcommand{\beq}{\begin{equation}}
\newcommand{\eeq}{\end{equation}}
\newcommand{\bea}{\begin{eqnarray}}
\newcommand{\eea}{\end{eqnarray}}
\newcommand{\be}{\begin{equation}}      
\newcommand{\ee}{\end{equation}}

\def\nn{\nonumber}

\usepackage{slashed}

\begin{document}

\title{The radial action for massive particles in spherically symmetric geometries: Exact
resummation at any PM order}

\author{Donato Bini$^{1}$,  Giorgio Di Russo$^{2}$}  
  \affiliation{
$^1$Istituto per le Applicazioni del Calcolo ``M. Picone,''\\ CNR, I-00185 Rome, Italy\\
$^2$School of Fundamental Physics and Mathematical Sciences, Hangzhou Institute for Advanced Study, UCAS, Hangzhou 310024, China\\
}

\date{\today}

\begin{abstract}
We compute the massive particles radial action   along hyperboliclike geodesics in various spherically symmetric spacetimes: the standard $4d$ Schwarzschild spacetime, its $d$-dimensional genralization known as Schwarzschild-Tangherlini solutions and for the D3-branes spacetimes, showing useful  resummation properties in terms of special (hypergeometric, Fox-Wright) functions in the eikonal limit and generalizing previous results valid for null geodesics. 
As a consequence, the scattering angle can be resummed too,  and we explicitly display the resummed expressions.

In addition, in the more interesting situation of a $4d$ Schwarzschild black hole spacetime, following the approach of the quantum Seiberg-Witten curves to the radial equation associated with a massive scalar field, we show that the quantum $a$-cycle (or, equivalently, the \lq\lq renormalized angular momentum") is simply related to radial action also in this massive case,  providing fully resummed expressions.
Finally, we display the explicit, expanded-form expression of the dual $a_D$-cycle, for which, however, no resummed expressions have been derived yet. 

\end{abstract}

\maketitle

\section{Introduction}
\label{Intro}

After more than one hundred years from its discovery in 1916, the Schwarzschild spacetime is still the preferred arena of many general relativistic calculations.
In this paper we are mainly concerned with the radial action for massive particles moving along  timelike, hyperboliclike geodesics of the equatorial plane of the Schwarzschild spacetime (as well as other spherically symmetric solutions).
The radial action, in fact, encodes most of the information necessary for the analytic treatment of the geodesics, and because of the Jeffrey-Wentzel-Kramer-Brilluoin  approximation, it is also a key tool to study perturbations  due to field of any spin-weight.

In the large literature existing for the Schwarzschild black hole solution, it is well known (and here is also shortly revisited in Appendix \ref{App_ellip}) how to write a closed form expression for the radial action in terms of elliptic functions, both for massless and massive particles. However, these exact expressions are poorly usable in the applications, and one has to resort to power series expansions.
For example, this is the case for hyperbolic-like orbits where the radial action is often presented in a large angular momentum expansion limit.

In  recent works \cite{Gonzo:2023goe,Ivanov:2025ozg,Parnachev:2020zbr} the radial action for null geodesics and hyperbolic-like motions, after being derived in the large angular momentum expansion limit, has been resummed in the eikonal limit in terms of hypergeometric functions (the large angular momentum expansion limit and the eikonal limit being closely related). This resummation property has also been tested (always in the massless case) in other spherically symmetric backgrounds, such as the topological star configuration \cite{Bini:2025ltr}, and even in axially symmetric solutions like the Kerr geometry \cite{Akpinar:2025huz,Bini:2025bll}.

Here, for the first time, we pass from the null orbit case to the that of massive particles, and show (see  Section \ref{rad_act_schw} below)  how to construct a resummed expression of the radial action for massive particles, in the eikonal limit. Consequently, the associated scattering angle can be given a (useful) resummed expression too. 

Another nice accomplishment of recent works \cite{Ivanov:2025ozg,Bini:2025ltr} is that of highlighting the link between the null particles radial action and the 
so-called \lq\lq renormalized angular momentum" $\nu$, entering in perturbation theory within the so called Mano-Suzuki-Takasugi (MST) approach,  when one considers a field of given spin (scalar, vectorial, tensorial) as perturbing the Schwarzschild spacetime \cite{Mano:1996vt,Sasaki:2003xr}.
This parameter actually takes  into account the monodromy property (at spatial infinity) of the solution of the radial equation, i.e.,  the nontrivial part of the  field after being separated in its variable dependence: from the angular variables by expansion in spin-weighted spherical harmonics, as well as from the time variable by Fourier transform (see Ref. \cite{Bini:2025ltr} for details). 

Here, we extend this link to the massive particles too, always working in the eikonal limit.

More precisely, in Section \ref{mass_scal_field}, we study massive scalar field waves on the Schwarzschild background. Unfortunately, the corresponding radial equation cannot be studied by using the MST formalism which is only available in the massless case, and hence one should use a different approach. 

Before proceeding, a few remarks are in order. Recently the MST approach has been extended to a particular 5D Einstein-Maxwell non-supersymmetric solution known as the Topological Star \cite{Bah:2020ogh,Bah:2020pdz}. These geometries admit a smooth and horizonless regime while preserving spherical symmetry and allowing for a separation between radial and angular dynamics. Owing to these properties, they serve as a toy model for more complex regular configurations such as fuzzballs. MST solutions have been used to compute the energy loss of a system in which a small massive particle moves and backreacts on the topological star geometry \cite{Bianchi:2024rod,DiRusso:2025lip,Bianchi:2025aei}. However, no attempts to consider massive fields within the MST formalism has been considered yet.

We use the results obtained long ago by Seiberg and Witten (SW) in the context of $\mathcal{N}=2$ Super Yang-Mills (SYM) theory with gauge group $SU(2)$, and in particular the instantonic calculus available in this gauge theory framework \cite{Seiberg:1994rs,Seiberg:1994aj}. This approach is equivalent to the MST formalism in the massless case.

A short summary of introductory details follows. 
\begin{enumerate}
\item In ${\mathcal N} = 2$ $SU(2)$ SYM theory, the Coulomb-branch moduli space can be classically\footnote{The term \lq\lq classical" is used here to distinguish this description from the \lq \lq quantum" one obtained through the embedding in the non-commutative Nekrasov-Shatashvili $\Omega$-background \cite{Gaiotto:2009we,Nekrasov:2009rc}; see below for details.} represented by an elliptic curve, which is a double cover of the complex plane with two branch cuts, and is topologically equivalent to a torus. Therefore, one can  define two different (dual) cycles, $a$ and $a_D$, forming a basis of its non-trivial homology.

\item SYM theories admit a low-energy effective description encoded in an analytic function known as the \lq\lq prepotential," denoted as ${\mathcal F}$ below.
In ${\mathcal N} = 2$ $SU(2)$ SYM, the prepotential is related, in a remarkable way, to the Coulomb-branch modulus parameter via the quantum Matone relation \cite{Matone:1995rx,Flume:2004rp}, i.e., to both $a$ and $a_D$. 
It is also worth to recall that, because of its structure, $a_D$ contains a tree-level and a one-loop contribution, together with an infinite series of instantonic corrections that can be computed perturbatively in power series of the gauge coupling. Due to non-renormalization theorems, no higher-loop contributions arise beyond one loop.

\item The above mentioned elliptic curve can be embedded into the Nekrasov-Shatashvili $\Omega$-background \cite{Nekrasov:2009rc}, and becomes a second order ordinary differential equation. On the other hand, in two-dimensional Liouville conformal field theory it is well known that the five-point conformal block with one degenerate insertion is governed by a Belavin-Polyakov-Zamolodchikov equation.  This equation  exhibits the same structure as the differential equations emerging from SW theory, namely, a Heun Equation (HE), i.e. a second-order ordinary differential equation with four regular singular points. This HE, or more precisely its confluent form, the Confluent Heun Equation (CHE), describes all perturbations of the Kerr-Newman  black hole family, including Schwarzschild. A vast, though not complete, collection of dictionaries between various gravity solutions and the corresponding SW curves and (confluent) forms of the HE equation is available in \cite{Bianchi:2022wku}. For recent developments on the solution of the CHE using the Floquet basis of solutions, see \cite{Fioravanti:2025bts}.

\item The Alday-Gaiotto-Tachikawa  duality establishes a correspondence between the partition functions of Liouville conformal field theory in two dimensions and the analogous object arising from $\mathcal{N}=2$ SYM theory in four dimensions with gauge group $SU(2)$ \cite{Alday:2009aq}.
Combining the two pictures, it is possible to derive formulae for the connection matrices relating the asymptotic expansions of the solutions near any two singularities of the HE. These matrices are given by the so called \lq \lq braiding" and \lq \lq fusion" rules of the underlying conformal field theory \cite{Bonelli:2022ten,Consoli:2022eey}.

\item
Finally, on the connection formulae it is  possible to impose the boundary conditions describing quasi normal modes (QNMs) frequencies. Given the solution with ingoing boundary conditions at the horizon, the connection formulae allow one to rewrite it as a linear combination of solutions ingoing and outgoing at infinity. The requirement that the coefficient of the ingoing wave at infinity vanishes gives an exact (and closed form) quantization condition for the QNMs. The latter quantization condition always contains the $a_D-$cycle \cite{Aminov:2020yma,Aminov:2023jve}.

\item This remarkable application of SW approach to black hole perturbation theory has been tested several times for different geometries, including both black holes and smooth, horizonless configurations such as fuzzballs \cite{Bianchi:2021xpr,Bianchi:2021mft,Bianchi:2022qph,Bianchi:2023sfs,DiRusso:2025qpf,Bianchi:2025ydq}.

The accuracy of QNM frequencies can be improved by including more and more instantonic contributions, which appear as a power series in the dimensionless frequency $M\omega$, with $M$ the mass of the black hole. As is clearly understandable, higher-order instantonic corrections are computationally expensive. In the present work, in the eikonal limit,  we show how to compute  high-order instantonic corrections, paving the way for future applications to high-precision analytic QNM calculations.  

\item Beyond QNM computations, the SW approach has also been used to compute tidal deformations \cite{DiRusso:2024hmd}, as well as the superradiant amplification factor \cite{Brito:2015oca,Cipriani:2024ygw}. More recently, it has also been applied to waveform reconstruction \cite{Bianchi:2024vmi,Cipriani:2025ikx,Cano:2026hlv}.
\end{enumerate}

To make a long story short,
by using the SW formalism  we  provide here a number of useful information for further developments of black hole perturbation theory.

Last, we enrich our discussion by considering the $d$-dimensional Schwarzschild-like solution (known as Schwarzschild-Tangherlini solution) as well as the case of the D3 branes spacetime, all solutions sharing the spherical symmetry. Noticeably, in the Schwarzschild-Tangherlini case we provide (partially) resummed expressions for either the radial action and the $d$-dimensional scattering angle involving the Fox-Wright functions (i.e., generalizations of the hypergeometric functions) which seem to be anchored to general $d$ dimension of the spacetime and are supposed to play a role when approaching regularization of divergent fields with dimensional regularization.

\section{Massive particles and fields in the Schwarzschild spacetime}

For convenience, we divide this section into several subsections.

\subsection{The radial action for massive particles in the Schwarzschild spacetime}
\label{rad_act_schw}
Let us consider the Schwarzschild spacetime with metric written in the standard form
\beq
ds^2=-f(r)dt^2+\frac{dr^2}{f(r)}+r^2(d\theta^2+\sin^2\theta d\phi^2)\,,
\eeq
where $f(r)=1-2M/r$.
The on-shell condition for a mass $\mu$ particle in motion on the equatorial plane of the Schwarzschild spacetime (without any loss of generality due to the spherical symmetry)  reads
\beq
g^{\mu\nu}P_\mu P_\nu +\mu^2=0\,,
\eeq
or, explicitly
\beq
-\frac{E^2}{f(r)}+P_r^2 f(r)+\frac{J^2}{r^2}+\mu^2=0\,.
\eeq

In terms of the dimensionless variables $u=M/r$, $E/\mu=\gamma=\sqrt{1+p_\infty^2}$, $P_\phi=J$, $J/(M\mu)=j$, $p_r=P_r/\mu$ (in units such that $G=c=1$; physical units will be restored later) this condition becomes
\beq
-\frac{\gamma^2}{1-2u}+p_r^2 (1-2u)+j^2u^2+1=0\,,
\eeq
from which
\beq
p_r^2=\frac{\gamma^2-(1-2u)(1+j^2u^2)}{(1-2u)^2}\,.
\eeq
The radial action for massive particles   in (equatorial) hyperbolic-like motion is given by
\beq
I_r^{\rm hyp}(\gamma,j) =\int_{r_0}^\infty p_r(\gamma,j; r) dr\,,
\eeq
where $r_0=r_0(\gamma,j)$ is the minimum approach distance, solution of the equation $p_r(\gamma,j; r)=0$, and
\bea
\label{integrand}
p_r dr &=& -\frac{\sqrt{-1 + \gamma^2 + 2 u - j^2 u^2 + 2 j^2 u^3}}{u^2 (1-2 u)}du\nonumber\\
&=& -\frac{j\sqrt{2(u-u_1)(u-u_2)(u-u_3)}}{u^2 (1-2 u)}du\,.
\eea
It is often convenient to introduce the   \lq\lq effective potential"   
\beq
V_{\rm eff}(u,j)=(1-2u)(1+j^2u^2)\,,
\eeq
such that 
\beq
p_r dr = -\frac{\sqrt{\gamma^2 -V_{\rm eff}(u,j)}}{u^2 (1-2 u)}du\,.
\eeq
Eq. \eqref{integrand} implies
\beq\label{Irint}
I_r^{\rm hyp}(\gamma,j)=j\int_0^{u_{\rm max}}\frac{\sqrt{2(u-u_1)(u-u_2)(u-u_3)}}{u^2 (1-2 u)}du\,.
\eeq
We  assume  (scattering regime) the following ordering of the roots is $u_1>u_3>0>u_2$ (besides the additional constraint $u_1+u_2+u_3=\frac12$) so that $u_{\rm max}=u_3$.
The three roots $u_1$, $u_2$, $u_3$ can be determined exactly
\bea
u_1(\gamma,j)&=& \frac16  + \frac{1}{6j}\left[ B^{1/3}  + \frac{(j^2-12)}{B^{1/3}}\right]\,,\nonumber\\
u_2(\gamma,j)&=& \frac16 + \frac{i}{6j}\left[ B^{1/3}e^{i\pi/6} -\frac{(j^2-12) e^{-i\pi/6}}{B^{1/3}}
\right]\,,\nonumber\\
u_3(\gamma,j)&=&\frac16 - \frac{i}{6j}\left[ B^{1/3}e^{-i\pi/6} -\frac{(j^2-12) e^{i\pi/6}}{B^{1/3}}
\right]\nonumber\\
&=&u_2(\gamma,-j) 
\,,
\eea
having denoted
\bea
A&=& \sqrt{48 + 3 (8 - 36 \gamma^2 + 27 \gamma^4) j^2 - 3 (\gamma^2-1) j^4}\,,\nonumber\\
B&=& 6  A + 18(2 - 3  \gamma^2)  j + j^3\,.
\eea
The radial action integral \eqref{Irint} admits an exact (even if formal) representation: 1) in terms of Elliptic integrals (which we discuss in Appendix \ref{App_ellip}  since all explicit expressions are much involved); 
 2) in terms of Lauricella hypergeometric functions \cite{Gonzo:2023goe}, which are defined as an Euler-type integral representation
\bea\label{Lau}
&&F_D^{(n)}(a,b_1,\dots,b_n,c;x_1,\dots,x_n)=\frac{\Gamma(c)}{\Gamma(a)\Gamma(c-a)}\times\nonumber\\
&&\int_0^1z^{a-1}(1-z)^{c-a-1}(1-x_1 z)^{-b_1}\dots(1-x_n z)^{-b_n}dz\,,\nonumber\\
\eea
or, equivalently, via the series representation
\bea
&&F_D^{(n)}(a,b_1,\dots,b_n,c;x_1,\dots,x_n)=\\
&&\sum_{k_1,\dots,k_n=0}^\infty 
\frac{(a)_{k_1+\cdots+k_n}\,(b_1)_{k_1}\cdots(b_n)_{k_n}}
{(c)_{k_1+\cdots+k_n}\;k_1!\cdots k_n!}\;
x_1^{k_1}\cdots x_n^{k_n}\,.\nonumber
\eea
Eq. \eqref{Irint}, rewritten by using $z=u/u_3$ as a new radial coordinate, needs an infrared regularization (introducing a dimensionless scale $\epsilon$ taken to be zero later) and can be cast into the form
\beq\label{regIr}
I_r^{\rm reg}=\lim_{\epsilon\to 0}  \int_0^1 dz {\mathcal I}_{r}^\epsilon(z)\,,
\eeq
where
\beq
{\mathcal I}_{r}^\epsilon(z)=C_{r,\epsilon} \frac{z^{\epsilon-2}}{(1-2u_3 z)}(1{-}z)^{\frac{1}{2}}\left(1-u_{3/1}z\right)^{\frac{1}{2}}\left(1-u_{3/2}z\right)^{\frac{1}{2}}\,,
\eeq
with 
\beq
u_{3/1}=\frac{u_3}{u_1}\,,\qquad u_{3/2}=\frac{u_3}{u_2}\,,
\eeq 
and 
\beq
C_{r,\epsilon}=j\frac{\sqrt{-2 u_1 u_2}}{u_3^{\frac{1}{2}-\epsilon}}\,.
\eeq
Consequently, using the definition \eqref{Lau}, Eq.  \eqref{regIr} becomes
\bea
I_r^{\rm reg} &=&\lim_{\epsilon \to 0} j \frac{\sqrt{-\pi u_1 u_2}}{\sqrt{2}u_3^{\frac{1}{2}-\epsilon}}
\frac{\,\Gamma(-1+\epsilon)}{\Gamma(\frac{1}{2}+\epsilon)}\times\\
&&F_D^{(3)}\!\left({-}1{+}\epsilon,{-}\frac{1}{2},{-}\frac{1}{2},1,\frac{1}{2}{+}\epsilon;\frac{u_3}{u_1},\frac{u_3}{u_2},2u_3\right)\,.\nonumber
\eea
In a large-$j$ expansion limit (at fixed $\gamma$) the roots $u_1$, $u_2$ and $u_3$ have the following expansions
\bea
u_1(\gamma,j)&=&\frac{1}{2}- \frac{ 2 \gamma^2 }{j^2} + \frac{8 \gamma^2(1 - 2\gamma^2)}{j^4}\nonumber\\
&&-\frac{ 32 \gamma^2(1 - 6 \gamma^2 + 7 \gamma^4)}{j^6} +O\left(\frac{1}{j^8}\right)\,,
\nonumber\\
u_2(\gamma,j)&=&- \frac{\sqrt{\gamma^2-1}}{j}+ \frac{\gamma^2}{j^2} +\frac{\gamma^2(4  - 5 \gamma^2)}{ 2 \sqrt{\gamma^2-1}j^3} 
+O\left(\frac{1}{j^4}\right)\,, \nonumber\\
u_3(\gamma,j)&=&\frac{\sqrt{\gamma^2-1}}{j} + \frac{\gamma^2}{j^2} 
-
\frac{ \gamma^2 (4 - 5 \gamma^2) }{ 
 2 \sqrt{\gamma^2-1} j^3 } 
+O\left(\frac{1}{j^4}\right)\nonumber\\
&=& u_2(\gamma, -j)\,,
\eea
with
\bea
u_{3/1}&=& \frac{2\sqrt{\gamma^2-1}}{j} 
+\frac{ 2\gamma^2}{ j^2} 
+ \frac{ \gamma^2 (-12 + 13 \gamma^2)}{ j^3\sqrt{\gamma^2-1}}+O\left(\frac{1}{j^4}\right)\,,\nonumber\\
u_{3/2}&=& -1  - \frac{2 \gamma^2}{j\sqrt{\gamma^2-1}} 
- \frac{ 2 \gamma^4 }{ j^2 (\gamma^2-1) }\nonumber\\
&+& \frac{ -8 \gamma^2 + 20 \gamma^4 - 13 \gamma^6 }{j^3 (\gamma^2-1)^{3/2}}+O\left(\frac{1}{j^4}\right)\,.
\eea
Unfortunately both these approaches, namely the elliptic integrals and Lauricella functions,  have the disadvantage that their expanded form should concern the expansion of both the arguments  and the parameters of the corresponding functions, implying long computational times as well as long expressions which, in the end, lead to merely formal results only.

Alternatively, giving up with the idea of having  exact expressions to deal with, one can compute $I_r^{\rm hyp}(\gamma,j)$ in a Post-Minkowskian (PM) sense, i.e.,  expanding in small $G$ (or large-$j$ or  small $u=GM/r$ expansion, with the product $ju$ fixed since the $G$ present in both of them cancels out) the integrand \eqref{integrand}. 
Passing to the new variable  $u=\frac{\sqrt{\gamma^2-1}}{j}\xi$, Eq. \eqref{integrand} becomes
\bea
\label{integrand3}
p_r dr &=& \frac{j^{3/2}\sqrt{jp_\infty (1-\xi^2)+2\xi(1+p_\infty^2\xi^2)}}{\sqrt{p_\infty} \xi^2 (j-2p_\infty \xi)} d\xi\,.\qquad
\eea
Note that the integration over $\xi$ should span the interval $[0,\xi_{\rm max}]$, where
\bea
\xi_{\rm max}&=& 1 +\frac{  \frac{1}{p_\infty} + p_\infty }{j} + \frac{ 3 + \frac{1}{2 p_\infty^2} + \frac{5 p_\infty^2}{2}}{j^2}\nonumber\\
&+& O\left(\frac{1}{j^3}\right)\,,
\eea
where
\beq
p_\infty=\sqrt{\gamma^2-1}\,,
\eeq
and, hence, will   formally involve an infinite number of contributions coming
from the expansion of $\xi_{\rm max}$. 
However,
 Ref. \cite{Damour:1988mr} has shown that the correct value of the PM-expanded integral can equally be obtained  by  using as upper limit
of the integral $\xi_{\rm max}=1$, 
PM-expanding the integrand taking the Hadamard
{\it Partie finie} 
(Pf) of all the divergent integrals generated in this way.

It is convenient to  introduce the notation
\beq
\label{x_definition}
\frac{x}{3\sqrt{3}}=\frac{p_\infty}{j}\equiv \frac{GM}{b}\equiv \frac{1}{\hat b},
\eeq
Moreover,  denoting
\beq
c=-\frac{2\xi(1+p_\infty^2\xi^2)}{p_\infty^2 (1-\xi^2)}\,,
\eeq
the integrand \eqref{integrand3} becomes
\begin{widetext}
\bea
\label{integrand4}
p_r dr &=& \hat b p_\infty \frac{\sqrt{1-\xi^2}}{\xi^2} \frac{\sqrt{1-\frac{c}{\hat b}}}{1-\frac{2\xi}{\hat b}}
 d\xi= -\hat b p_\infty \frac{\sqrt{1-\xi^2}}{\xi^2} \sum_{n=0}^\infty \frac{\left(n-\frac{3}{2}\right)!c^n }{2\sqrt{2}n!\hat{b}^n}{}_2F_1\left(1,-n,\frac{3}{2}-n,\frac{2\xi}{a}\right)d\xi\nonumber\\
 &=& -\frac{\hat{b}p_{\infty}\sqrt{1-\xi^2}}{\xi^2}d\xi\sum_{{n,s}=0}^\infty\frac{c^n}{2\sqrt{2}\hat{b}^n}\frac{\Gamma\left(n-\frac{1}{2}\right)\Gamma(1+s)\Gamma(s-n)\Gamma\left(\frac{3}{2}-n\right)}{\Gamma(n+1)\Gamma(-n)\Gamma(\frac{3}{2}-n+s)\Gamma(s+1)}\left(\frac{2\xi}{c}\right)^s\nonumber\\
&=&\sum_{n,s,k=0}^\infty\frac{2^{n-\frac{3}{2}} \hat{b}^{1-n} (-1)^{s-n+1} \tan (\pi  n) p_{\infty
   }^{1-2 k}  \Gamma (s-n)
   \binom{n-s}{k} }{\Gamma
   \left(-n+s+\frac{3}{2}\right)}\xi ^{-2 k+3 n-2 s-2}\left(1-\xi ^2\right)^{\frac{1}{2}-n+s}d\xi\,.
\eea
Therefore, recalling that
\beq
\int_0^1 d\xi  (1-\xi^2)^{a}\xi^{b}=\frac{\Gamma(1 + a) \Gamma[\frac{1 + b}{2}]}{2 \Gamma[\frac32 + a + \frac{b}2]}
\eeq
we can  perform the integration over $\xi$ obtaining
\bea
\label{seires_Ir}
I_r&=&\sum_{n,s,k=0}^\infty\frac{2^{n-\frac{5}{2}}
   \hat{b}^{1-n} (-1)^{s-n+1} \tan
   (\pi  n) p_{\infty }^{1-2 k}
   \Gamma (s-n) \binom{n-s}{k}
   \Gamma \left(-k+\frac{3
   n}{2}-s-\frac{1}{2}\right)}{\Gamma
   \left(-k+\frac{n}{2}+1\right)}\,,
\eea
\end{widetext}
where specific values of $n$ (like $n=s=k=0$) may imply the vanishing or the divergence of the various terms in sum, and therefore are meant to be taken as a limit.
The nice feature of Eq. \eqref{seires_Ir} is that the three series involved can be resummed, leading to hypergeometric functions, and hence to \lq\lq almost exact" expressions, of direct and simple use in the applications. 

Let us show this in detail.
The PM-expanded final result
can be split in a $\pi$-part ($p_\infty$-even) and a non-$\pi$-part (or ${\slashed{\pi}}$, $p_\infty$-odd),
\beq
I_r^{\rm hyp}=I^{\rm hyp}_{r\, \pi}+I^{\rm hyp}_{r\, {\slashed{\pi}}}\,,
\eeq
with
\bea
I^{\rm hyp}_{r\,  \pi }&=&\pi \sum_{k=0}^\infty\frac{{\mathcal I}_{2k-1}^{ \pi }}{j^{2k-1}}\,,\nonumber\\
I^{\rm hyp}_{r\, {\slashed{\pi}}}&=&{\mathcal I}_{0}^{\slashed{\pi}}\ln j+\sum_{k=1}^\infty\frac{{\mathcal I}_{2k}^{\slashed{\pi}}}{j^{2k}}\,,
\eea
with
\bea\label{Irpi}
{\mathcal I}_{-1}^{ \pi } &=& -\frac12\,,  \nonumber\\
{\mathcal I}_1^{ \pi } &=&  \frac32+  \frac{ 15}{8} p_\infty^2\,, \nonumber\\
{\mathcal I}_3^{ \pi }&=&   \frac{35 }{8} + \frac{105 p_\infty^2}{8}+ \frac{1155 p_\infty^4}{128}\,, \nonumber\\
{\mathcal I}_5^{ \pi }&=&   \frac{231}{8 }+ \frac{9009p_\infty^2}{64}+ \frac{27027p_\infty^4}{128} + \frac{51051p_\infty^6}{512} \,,\nonumber\\
\eea
and
\begin{equation*}
\label{I_non_pi_defs}
    {\mathcal I}_0^{\slashed{\pi}} =\left(\frac{1}{p_\infty}+2 p_\infty\right)  \,, \nonumber\\ 
\end{equation*}
\bea\label{Irpi2}
{\mathcal I}_2^{\slashed{\pi}}&=& -\frac{1}{6 p_\infty^3} + \frac{2}{p_\infty} + 12 p_\infty + \frac{32 p_\infty^3}{3}\,,  \nonumber\\
{\mathcal I}_4^{\slashed{\pi}}&=&  \frac{1}{20 p_\infty^5} - \frac{1}{2 p_\infty^3} + \frac{8}{p_\infty} + 80 p_\infty + 160p_\infty^3 + \frac{448 p_\infty^5}{5} \,,\nonumber\\
{\mathcal I}_6^{\slashed{\pi}}&=&  -\frac{1}{42 p_\infty^7} + \frac{4}{15 p_\infty^5} - \frac{8}{3 p_\infty^3} + \frac{160}{3 p_\infty} 
+\frac{
 2240 p_\infty}{3}\nonumber\\ 
&+& \frac{7168 p_\infty^3}{3} + \frac{14336 p_\infty^5}{5} + \frac{
 8192 p_\infty^7}{7} \,,
\eea
to list only  the first terms.
Note that ${\mathcal I}_k^{ \pi }$ contains terms starting from $p_\infty^0$ up to $p_\infty^{k+1}$ (with a step of $p_\infty^2$), while $ {\mathcal I}_k^{\slashed{\pi}}$ contains terms starting as $\frac{1}{p_\infty^{k+1}}$ and going up to $p_\infty^{k+1}$ (with a step of $p_\infty^2$).

Our accomplishment is the following
\beq
I_r=I_r^0+\mu M\sum_{k=0}^\infty p_\infty^{1-2k} W(k)\,,
\eeq
where
\beq\label{Ir0d4}
I_r^0={\mathcal I}_0^{\slashed{\pi}}\ln j=\left(\frac{1}{p_\infty}+2 p_\infty\right)\ln (j)\,,
\eeq
and
\bea
W(k) &=&  C_1(k)x^2 \mathcal{F}_1 +C_2(k) \frac{1}{x}\mathcal{F}_2\,,
\eea
with
\begin{widetext}
\bea
C_1(k)&=&  2^{k-3} 3^{-k-\frac12}
 \frac{ (k - 3)\Gamma\left( \frac{5}{3} - \frac{k}{3} \right) \Gamma\left( \frac{7}{3} - \frac{k}{3} \right) \Gamma\left( -\frac{k}{3} \right)  }{ \Gamma(k) }\,,   \nonumber\\
C_2(k)&=& (-1)^{k+1} 2^{3k-1} 3^{3/2} 
 \frac{\pi^2   }{ \Gamma(2k + 1) }\,, \nonumber\\
{\mathcal F}_1&=& _5\tilde{F}_4\left(1,1,\frac{5}{3}-\frac{k}{3},2-\frac{k}{3},\frac{7}{3}-\frac{
   k}{3};2,\frac{5}{2}-k,2-\frac{k}{2},\frac{5}{2}-\frac{k}{2};x^2\right)\,,\nonumber\\
{\mathcal F}_2&=& _5\tilde{F}_4\left(-\frac{1}{2},1,\frac{1}{6}-\frac{k}{3},\frac{1}{2}-\frac{k}{
   3},\frac{5}{6}-\frac{k}{3};\frac{1}{2},1-k,\frac{1}{2}-\frac{k}{2},1-\frac{k}{2
   };x^2\right)\,,
\eea
and ${}_p\tilde F_q$ the regularized hypergeometric functions. Noticeably, the expansion parameter,  (i.e., impact parameter like variable $x$) only enters the argument of the hypergeometric functions and not the coefficients.
\end{widetext} 
For example, for $k=0$
\bea
C_1(0)&=&  -2^{-3} 3^{\frac12}\Gamma\left( \frac{5}{3}\right) \Gamma\left( \frac{7}{3}\right)\lim_{k\to 0}\frac{ \Gamma\left( -\frac{k}{3} \right)}{\Gamma(k)} =\frac{2\pi}{9}\,, \nonumber\\
C_2(0)&=& - 2^{-1} 3^{3/2} \frac{\pi^2   }{ \Gamma(1) }=- \frac{3\sqrt{3}  \pi^2}{2} \,, \nonumber\\
{\mathcal F}_1&=& _5\tilde{F}_4\left(1,1,\frac{5}{3},2,\frac{7}{3};2,\frac{5}{2},2,\frac{5}{2};x^2\right)\nonumber\\
&=& \frac{16}{9\pi}{}_4F_3 \left(1,1,\frac53,\frac73; 2, \frac52,\frac52; x^2\right)\,,\nonumber\\
{\mathcal F}_2&=& _5\tilde{F}_4\left(-\frac{1}{2},1,\frac{1}{6},\frac{1}{2},\frac{5}{6};\frac{1}{2},1,\frac{1}{2},1;x^2\right)\nonumber\\
&=& \frac{1}{\pi} {}_3F_2\left(-\frac12,\frac16,\frac56; \frac12, 1; x^2\right)\,,
\eea
so that
\bea
W(0)&=&\frac{32 }{81}  x^2\, {}_4F_3 \left(1,1,\frac53,\frac73; 2, \frac52,\frac52; x^2\right) \nonumber\\
&-& \frac{3\sqrt{3}  \pi}{2}\frac{1}{x} {}_3F_2\left(-\frac12,\frac16,\frac56; \frac12, 1; x^2\right)\,.
\eea
Similarly,
\bea
W(1)&=&\frac{4}{9} x^2   {}_4F_3\left(1,1, \frac43, \frac53; \frac32,\frac32, 2; x^2\right)\nonumber\\
&+&\frac{\pi x}{2\sqrt{3}} \, {}_3F_2 \left(\frac12,\frac56,\frac76; 1, \frac32; x^2\right) \,,
 \nonumber\\
W(2)&=&\frac{2}{27} x^2   {}_4F_3\left(1,1, \frac43, \frac53; \frac12,\frac32, 2; x^2\right)\nonumber\\
&+&\frac{35 \pi x^3}{648\sqrt{3}} \, {}_3F_2 \left(\frac32,\frac{11}6,\frac{13}6; 2, \frac52; x^2\right) 
 \,,\nonumber\\
W(3)&=&-\frac{1}{162} x^2   {}_4F_3\left(1,1,\frac23, \frac43; -\frac12,\frac12, 2; x^2\right)\nonumber\\
&+&\frac{77 \pi x^5}{5832\sqrt{3}} \, {}_3F_2 \left(\frac52,\frac{13}6,\frac{17}6; 2, \frac72; x^2\right) 
 \,, \nonumber\\
W(4)&=& \frac{5}{192} \Big[   {}_3F_2\left(-\frac23,-\frac13,1; -\frac52,-\frac12; x^2\right)-1\Big]\nonumber\\
&-&\frac{1}{216} x^2   {}_4F_3\left(1,1,\frac13, \frac23; -\frac32,\frac12, 2; x^2\right)\nonumber\\
&+& \frac{3575 \pi x^7}{839808\sqrt{3}} \, {}_4F_3 \left(\frac72,\frac72,\frac{17}6,\frac{19}6; \frac52, 3, \frac92; x^2\right) 
\,, \nonumber\\
\eea
leading to
\bea
\frac{I_r}{M\mu}&=& I_r^0+ p_\infty W(0)+\frac{1}{p_\infty} W(1)+\frac{1}{p_\infty^3} W(2)\nonumber\\
&+& \frac{1}{p_\infty^5} W(3)+\frac{1}{p_\infty^7} W(4)\ldots
\eea
Recalling the definition \eqref{x_definition}, differentiation with respect to $j$ of the radial action leads to the scattering angle
\bea
\frac{\chi+\pi}{2}&=& \frac{\chi_1}{j}+\sum_{k=0}^\infty p_\infty^{-2k}\, H(k)\,,
\eea
where
\bea
H(k)&=&  D_1 x^3 {\mathcal H}_1 + D_2 {\mathcal H}_2\,, 
\eea
with
\begin{widetext}
\bea
D_1
&=& 2\cdot 3^{-3/2}  C_1\,, \nonumber\\
D_2
&=&  -  \frac{C_2}{ 3^{3/2}\sqrt{\pi}}\,,\nonumber\\
{\mathcal H}_1&=& _4\tilde{F}_3\left(1,\frac{5-k}{3},\frac{6-k}{3},\frac{7-k}{3};\frac{5}{2}-k,2-\frac{k}{2},\frac{5-k}{2};x^2\right)\,, \nonumber\\
{\mathcal H}_2&=& _4\tilde{F}_3\left(1,\frac{1-2k}{6},\frac{3-2k}{6},\frac{5-2k}{6};1-k,\frac{1-k}{2},\frac{2-k}{2};x^2\right)
\,.\nonumber\\
\eea
For example,
\bea
H(0)&=& \frac{64 x^3 \, _3F_2\left(1,\frac{5}{3},\frac{7}{3};\frac{5}{2},\frac{5}{2};x^2\right)}{243 \sqrt{3}}+\frac{1}{2} \pi  \, _2F_1\left(\frac{1}{6},\frac{5}{6};1;x^2\right)\,,\nonumber\\
H(1)&=& \frac{1}{162} x^2 \left(16 \sqrt{3} x \, _3F_2\left(1,\frac{4}{3},\frac{5}{3};\frac{3}{2},\frac{3}{2};x^2\right)+9 \pi  \, _2F_1\left(\frac{5}{6},\frac{7}{6};1;x^2\right)\right)\,,\nonumber\\
H(2)&=& \frac{x^3 \left(32 \sqrt{3} \, _3F_2\left(1,\frac{4}{3},\frac{5}{3};\frac{1}{2},\frac{3}{2};x^2\right)+35 \pi  x \, _2F_1\left(\frac{11}{6},\frac{13}{6};2;x^2\right)\right)}{1944}\,,\nonumber\\
H(3)&=& \frac{385 \pi  x^6 \, _2F_1\left(\frac{13}{6},\frac{17}{6};2;x^2\right)-72 \sqrt{3} x^3 \, _3F_2\left(\frac{2}{3},1,\frac{4}{3};-\frac{1}{2},\frac{1}{2};x^2\right)}{52488}\,,\nonumber\\
H(4)&=& \frac{25025 \pi  x^8 \, _3F_2\left(\frac{17}{6},\frac{19}{6},\frac{7}{2};\frac{5}{2},3;x^2\right)-2304 \sqrt{3} x^5 \,
   _3F_2\left(\frac{4}{3},\frac{5}{3},2;-\frac{1}{2},\frac{3}{2};x^2\right)}{7558272}\,.
\eea
\end{widetext}

Taking these first five terms above, expanding in powers of $x$ and then restoring (in order to well known expressions) $p_\infty$ and $j$ from Eq. \eqref{x_definition},
we find
\bea
\frac{\chi+\pi}{2}&=& \frac{\pi}{2}+
\frac{\frac{1}{p_\infty}+2 p_\infty}{j}
+\frac{\frac{15 \pi p_\infty^2}{8}+\frac{3 \pi }{2}}{j^2}\nonumber\\
&+&\frac{\frac{64p_\infty^3}{3}-\frac{1}{3p_\infty^3}+24p_\infty+\frac{4}{p_\infty}}{j^3}\nonumber\\
&+&\frac{\frac{3465 \pi p_\infty^4}{128}+\frac{315 \pi p_\infty^2}{8}+\frac{105 \pi }{8}}{j^4}\nonumber\\
&+&\frac{\frac{1792p_\infty^5}{5}+640p_\infty^3-\frac{2}{p_\infty^3}+320p_\infty+\frac{32}{p_\infty}}{j^5}\nonumber\\
&+&\frac{\frac{255255 \pi p_\infty^6}{512}+\frac{135135 \pi p_\infty^4}{128}+\frac{45045 \pi p_\infty^2}{64}+\frac{1155 \pi }{8}}{j^6}\nonumber\\
&+&\frac{\frac{49152p_\infty^7}{7}+\frac{86016p_\infty^5}{5}+14336p_\infty^3+4480p_\infty+\frac{320}{p_\infty}}{j^7}\nonumber\\
&+&\frac{1}{j^8}\left( \frac{334639305 \pi p_\infty^8}{32768}+\frac{14549535 \pi p_\infty^6}{512}\right.\nonumber\\
&+&\left.\frac{14549535 \pi p_\infty^4}{512}+\frac{765765 \pi p_\infty^2}{64}+\frac{225225 \pi}{128}\right)\nonumber\\
&+& O\left(\frac{1}{j^9}\right)\,,
\eea
which reproduces known results.

\subsection{Massive Scalar waves in  Schwarzschild}
\label{mass_scal_field}

The massive scalar wave equation in the Schwarzschild spacetime is given  by  
\beq
\Box\Phi=\frac{\mu^2}{\hbar^2}\Phi\,,
\eeq
and can be studied by using separation of variables, i.e., by expanding the angular part in (scalar) spherical harmonics and Fourier transforming the time variable
\beq
\Phi(t,r,\theta,\phi)=\sum_{\ell m}Y_{\ell m}(\theta,\phi)\int \frac{d\omega}{2\pi}e^{-i\omega t} R_{lm\omega}(r)\,.
\eeq
Here we use units are such that $\hbar\not =1$, to facilitate taking the eikonal limit later.
The radial motion  is described by the following Heun-type equation, 
\bea
&&r(r-2M)R_{\ell m\omega}''(r)+2(r-M)R_{\ell m\omega}'(r)\nonumber\\
&&\qquad +\Big[\frac{\omega^2 r^3}{r-2M}-\ell(\ell{+}1)-\frac{\mu^2 r^2}{\hbar^2}\Big]R_{\ell m\omega}(r){=}0\,.\qquad
\eea
The latter can be cast in its normal form  via the rescaling
\beq
R_{\ell m\omega}(r)=\frac{\psi_{\ell m\omega}(r)}{r\sqrt{1-\frac{2M}{r}}}\,,
\eeq
so that it becomes
\beq
\psi_{\ell m\omega}''(r)+Q_{\rm W}(r)\psi_{\ell m\omega}(r)=0\,,
\eeq
with  
\beq\label{QW}
Q_{\rm W}(r){=}\frac{\omega^2r^4{-}\ell(\ell{+}1)r(r{-}2M){+}M^2}{r^2(r{-}2M)^2}{-}\frac{r\mu^2}{\hbar^2(r{-}2M)}\,.
\eeq

To solve this type of equation,  in the analogous massless case only, Mano-Suzuki-Takasugi (MST) have developed a formalism which is largely used in black hole perturbation theory and it is particularly efficient when working in Post-Newtonian sense.
Unfortunately, the same formalism has not been developed yet in the massive case, i.e., the one under consideration here.
One must then resort an equivalent approach, less familiar in the classical general relativity community,  based on the quantum Seiberg-Witten (qSW), which instead can be easily adapted to the massive case too.
Actually, one looks for solutions which are purely in-going at the horizon (in-solutions) and purely out-going at infinity (up-solutions). The latter can be represented as an infinite series of hypergeometric functions. Their \lq\lq monodromy property"  at infinity is characterized by the so called renormalized angular momentum, $\nu$, which is a generalization of the (quantum) angular momentum number $\ell$ and enters both the in- and up-solutions..

The radial action, the parameter $\nu$ and the $a$-cycle in the massless case and in the eikonal limit are simply related, or, equivalently the qSW and the MST approaches are simply related, the relation being a gauge-gravity  dictionary. 
Indeed from the MST point of view, recent works in the massless case have shown that $\nu$ is simply related to the radial action of geodesics in the eikonal limit, while, from the qSW point of view, $\nu$ has been related to the so-called $a$-cycle appearing in the quantum version of the Seiberg-Witten curves \cite{Bianchi:2024vmi,Bini:2025ltr,Bini:2025bll}.  

As we will show below, this link is still valid in the massive case, which is another accomplishment of the present work.

\subsection{The quantum Seiberg-Witten curves}

Following Ref. \cite{Bianchi:2021mft} and in order to get some intuition on the gauge-gravity dictionary  let us very briefly recall the \lq \lq classical" Seiberg-Witten (SW) curve for an $SU(2)$ gauge theory, with $N_f$ hypermultiplets with masses $m_i$ in flat space \cite{Witten:1997sc}. 
The SW curve describes the dynamics of the $SU(2)$ gauge theory, and can be written in the form of an elliptic curve
\begin{equation}\label{classcurve}
q y^2P_L(x)+y P_0(x)+P_R(x)=0\,,
\end{equation}
with
\bea\label{PRL0}
P_R(x)&=&\prod_{i=1}^{N_R}(x-m_i)\,,\nonumber\\ 
P_L(x)&=&\prod_{i=3}^{2+N_L}(x-m_i)\,,\nonumber\\ 
P_0(x)&=&x^2-u+q p_0(x)\,,
\eea
where $q$ is the gauge coupling, $u$ is the Coulomb branch modulus, and $p_0(x)$ a quadratic polynomial in the variable $x$. 
The  function $p_0(x)$ as well as the numbers $N_R$ and $N_L$ (which together give the number of flavours, $N_f=N_R+N_L$) will be specified below 
for the relevant case under exam in this paper. 

The classical curve \eqref{classcurve} can be transformed into a Heun-type ordinary differential equation after embedding it in the non-commutative Nekrasov-Shatashvili background \cite{Nekrasov:2009rc}, which amounts to promoting the variables to operators obeying the following commutation relation
\begin{equation}
\label{commrel}
\Big[\hat{x},\ln \hat{y}\Big]=\hbar\,.
\end{equation}
The quantum curve follows from Eq. \eqref{classcurve}
\begin{equation}\label{ODESW}
    \Big[q \hat{y}^{\frac{1}{2}}P_L(\hat{x})\hat{y}^{\frac{1}{2}}+P_0(\hat{x})+\hat{y}^{-\frac{1}{2}}P_R(\hat{x})\hat{y}^{-\frac{1}{2}}\Big]U=0\,,
\end{equation}
where, for theories with less than four flavors,
\begin{equation}
\label{quantP0}
P_0(x)=x^2-u+q\delta_{N_f, 3}\left(x-\sum_i m_i+\frac{\hbar}{2}\right)+q\delta_{N_f,2}\,.
\end{equation}
Using the commutation relation \eqref{commrel}, one can view \eqref{ODESW} as an ordinary differential equation in the $y$-variable
\bea
\label{ODESWnew}
  &&  \Big[q y^2P_L\left(\hat{x}+\frac{\hbar}{2}\right)+y P_0(\hat{x})+P_R\left(\hat{x}-\frac{\hbar}{2}\right)\Big]U(y)
\nonumber\\
&&=\Big[A(y)\hat{x}^2+B(y)\hat{x}+C(y)\Big]U(y)=0\,.
\eea
For theories with three masses $m_1$, $m_2$ and $m_3$ one finds
\bea
A&=&1+y,\nonumber\\ 
B&=& qy^2+qy-m_1-m_2-\hbar\,,\nonumber\\
C&=&-qy^2\left(m_3-\frac{\hbar}{2}\right)-u y+q y\Big[\frac{\hbar}{2}-\sum_{i=1}^3 m_i\Big]\nonumber\\
&+&\left(m_1+\frac{\hbar}{2}\right)\left(m_2+\frac{\hbar}{2}\right)\,.
\eea
Eq.\eqref{ODESWnew} can be cast in its normal  form via the transformation 
\be
U(y)=\frac{1}{\sqrt{y}}e^{-\frac{1}{2\hbar}\int^y\frac{B(y')}{y'A(y')}dy'}\Psi(y)\,,
\ee
so that  
\be
\Psi''(y)+Q_{\rm SW}(y) \Psi(y)=0\,,
\ee
with
\be
Q_{\rm SW}(y)=\frac{4 AC -B^2+2\hbar y(B A'-A B')+\hbar^2A^2}{4\hbar^2 y^2 A^2}\,.
\ee

In the specific case  $N_f=3$ ($N_R=2$ and $N_L=1$) we have the quantum Seiberg-Witten   curve
\bea
\label{Q21SW}
Q_{2,1}&=&-\frac{q^2}{4\hbar^2}+\frac{\hbar^2-(m_1-m_2)^2}{4\hbar^2 y^2}+\frac{\hbar^2-(m_1+m_2)^2}{4\hbar^2(1+y)^2}\nonumber\\
&{-}&\frac{m_3 q}{\hbar^2 y}{+}\frac{{-}\hbar^2{+}2\hbar q{+}2(m_1^2{+}m_2^2{-}(m_1{+}m_2)q{-}2u)}{4\hbar^2 y(1+y)}\,.\nonumber\\
\eea
Using the commutation relation \eqref{commrel}, the form of the qSW curve in \eqref{ODESW} becomes
\be\label{ODESW2}
\Big[q P_L\left(x-\frac{\hbar}{2}\right)\hat{y}+P_0(x)+P_R\left(x+\frac{\hbar}{2}\right)\hat{y}^{-1}\Big]\tilde{U}(x)=0\,.
\ee
The quantum periods can be computed by solving the difference equation following from the qSW curve after setting $\hat{y}=e^{-\hbar\partial_x}$ and introducing the functions
\bea
W(x)&=&\frac{1}{P_R\left(x+\frac{\hbar}{2}\right)}\frac{\tilde{U}(x)}{\tilde{U}\left(x+\hbar\right)}\,,\nonumber\\
M(x)&=&P_L\left(x-\frac{\hbar}{2}\right)P_R\left(x-\frac{\hbar}{2}\right)\,.
\eea
Eq. \eqref{ODESW2} becomes then
\be
qM(x)W(x)W(x-\hbar)+P_0(x)W(x)+1=0\,,
\ee
and, following \cite{DiRusso:2024hmd}, together with its shifted form $x\to x+\hbar$, \eqref{ODESW2} becomes
\bea
\label{ODEacycl}
\frac{q M(a{+}\hbar)}{P_0(a{+}\hbar){-}\frac{q M(a{+}2\hbar)}{P_0(a{+}2\hbar){-}\dots}}{+}\frac{q M(a)}{P(a{-}\hbar){-}\frac{q M(a{-}\hbar)}{P_0(a{-}2\hbar){-}\dots}}{-}P_0(a){=}0.\nonumber\\
\eea
Eq. \eqref{ODEacycl} can be easily solved for $a$ order by order in $q$. 
The starting point of the recursion is the free theory $q=0$, where $a\sim\sqrt{u}$ and then the expression of $a$ is of the type
\be\label{aii}
a=\sqrt{u}+\sum_i a_i q^i\,.
\ee
The first three coefficients $a_i$ are shown below in Table \ref{ai}.

\begin{table*}  
\caption{\label{ai} List of the first three coefficients $a_i$ of the $a$-cycle: $a=\sqrt{u}+\sum_i a_i q^i$\,.
 }
\begin{ruledtabular}
\begin{tabular}{ll}
$a_1$&$\frac{4 \left(m_2+m_3\right) u+4 m_1 \left(m_2
   m_3+u\right)-\left(m_1+m_2+m_3\right) \hbar
   ^2-4 u \hbar +\hbar ^3}{4 \sqrt{u} \left(4
   u-\hbar ^2\right)}$\\
$a_2$& $\frac{1}{256 u^{3/2} \left(u-\hbar ^2\right) \left(4
   u-\hbar ^2\right)^3}\Big[16 m_1 \left(\hbar-m_2-m_3 \right) \left(4
   u^2-5 u \hbar ^2+\hbar ^4\right) \left(4 m_2
   m_3 \left(12 u-\hbar ^2\right)+\left(\hbar
   ^2-4 u\right)^2\right)$\\
   &$+\left(\hbar ^2-4
   u\right)^2 \left(m_2^2 \left(48 m_3^2 u-48
   u^2+44 u \hbar ^2-8 \hbar ^4\right)+16 m_2
   \left(\hbar -m_3\right) \left(4 u^2-5 u \hbar
   ^2+\hbar ^4\right)\right.$\\
   &$-\left.\left(4 u-\hbar ^2\right)
   \left(4 m_3 \left(m_3 \left(3 u-2 \hbar
   ^2\right)-4 u \hbar +4 \hbar ^3\right)+4
   u^2+3 u \hbar ^2-8 \hbar ^4\right)\right)+4
   m_1^2 \left(4 \left(m_2^2 \left(3 u
   \left(\hbar ^2-4 u\right)^2\right.\right.\right.$\\
   &$-\left.\left.\left.4 m_3^2 \left(60
   u^2-35 u \hbar ^2+2 \hbar ^4\right)\right)+4
   m_3 m_2 \left(-48 u^3+64 u^2 \hbar ^2-17 u
   \hbar ^4+\hbar ^6\right)+3 m_3^2 u
   \left(\hbar ^2-4 u\right)^2\right)\right.$\\
   &$-\left.\left(3
   u-2 \hbar ^2\right) \left(4 u-\hbar
   ^2\right)^3\right)\Big]\,,$\\
$a_3$& $\frac{1}{1024 u^{5/2} \left(4 u-9 \hbar ^2\right) \left(u-\hbar ^2\right)^2 \left(4 u-\hbar ^2\right)^5}\Big[-\left(\left(4 u-9 \hbar ^2\right) \left(\hbar -m_2-m_3\right) \left(\left(8 \hbar ^6-11 u \hbar ^4-3 u^2 \hbar
   ^2+4 u^3\right.\right.\right.$\\
   &$+\left.\left.\left.4 m_2 \left(\left(2 \hbar ^4-5 u \hbar ^2+5 u^2\right) m_2-4 \hbar  \left(u-\hbar
   ^2\right)^2\right)\right) \left(\hbar ^2-4 u\right)^2+4 \left(\left(\hbar ^2-4 u\right)^2 \left(2 \hbar ^4-5
   u \hbar ^2+5 u^2\right)\right.\right.\right.$\\
   &$-\left.\left.\left.12 u \left(\hbar ^4-15 u \hbar ^2+20 u^2\right) m_2^2\right) m_3^2-16 \left(\hbar
   ^4-5 u \hbar ^2+4 u^2\right)^2 \left(\hbar -m_2\right) m_3\right) \left(4 u-\hbar
   ^2\right)^3\right)$\\
   &$+ \left(\hbar ^2-4 u\right)^2 m_1 \left(\left(4 u-9 \hbar ^2\right) \left(24 \hbar ^6-43 u
   \hbar ^4+13 u^2 \hbar ^2-48 \left(u-\hbar ^2\right)^2 m_2 \hbar +4 u^3\right.\right.$\\
   &$+\left.\left. 4 \left(6 \hbar ^4-13 u \hbar ^2+9
   u^2\right) m_2^2\right) \left(4 u-\hbar ^2\right)^3+16 m_2 \left(-54 \hbar ^{10}+861 u \hbar ^8-6397 u^2
   \hbar ^6+13100 u^3 \hbar ^4\right.\right.$\\
   &$-\left.\left.10480 u^4 \hbar ^2+2880 u^5+4 u \left(27 \hbar ^6-497 u \hbar ^4+1120 u^2 \hbar
   ^2-560 u^3\right) m_2^2\right) m_3^3\right.$\\
   &$+\left.4 \left(4 u-9 \hbar ^2\right) \left(\left(6 \hbar ^4-13 u \hbar ^2+9
   u^2\right) \left(4 u-\hbar ^2\right)^3+4 m_2 \left(\left(12 \hbar ^8-181 u \hbar ^6+1235 u^2 \hbar ^4-1840
   u^3 \hbar ^2+720 u^4\right) m_2\right.\right.\right.$\\
   &$-\left.\left.\left.4 \hbar  \left(u-\hbar ^2\right)^2 \left(3 \hbar ^4-40 u \hbar ^2+240
   u^2\right)\right)\right) m_3^2-4 \left(12 \hbar  \left(u-\hbar ^2\right)^2 \left(9 \hbar ^2-4 u\right)
   \left(\hbar ^2-4 u\right)^3\right.\right.$\\
   &$+\left.\left.m_2 \left(108 \hbar ^{12}-1803 u \hbar ^{10}+15389 u^2 \hbar ^8-22296 u^3 \hbar
   ^6+2496 u^4 \hbar ^4+8320 u^5 \hbar ^2-2304 u^6\right.\right.\right.$\\
   &$+\left.\left.\left.4 m_2 \left(4 \hbar  \left(4 u-9 \hbar ^2\right) \left(3
   \hbar ^4-40 u \hbar ^2+240 u^2\right) \left(u-\hbar ^2\right)^2+\left(54 \hbar ^{10}-861 u \hbar ^8+6397 u^2
   \hbar ^6\right.\right.\right.\right.\right.$\\
   &$-\left.\left.\left.\left.\left.13100 u^3 \hbar ^4+10480 u^4 \hbar ^2-2880 u^5\right) m_2\right)\right)\right) m_3\right)-4 \left(9
   \hbar ^4-40 u \hbar ^2+16 u^2\right) m_1^2 \left(\hbar -m_2-m_3\right)$\\
   &$\times \left(\left(6 \hbar ^4-13 u \hbar ^2+9
   u^2\right) \left(\hbar ^2-4 u\right)^4-12 u \left(\hbar ^4-15 u \hbar ^2+20 u^2\right) m_3^2 \left(\hbar ^2-4
   u\right)^2\right.$\\
   &$+\left.16 \left(u{-}\hbar ^2\right)^2 \left(4 u{-}\hbar ^2\right) \left(3 \hbar ^4{-}40 u \hbar ^2{+}240
   u^2\right) m_2 m_3{+}12 m_2^2 \left(4 \left(2 \hbar ^8{-}39 u \hbar ^6{+}371 u^2 \hbar ^4{-}840 u^3 \hbar ^2{+}560
   u^4\right) m_3^2\right.\right.$\\
   &$-\left.\left.u \left(\hbar ^2-4 u\right)^2 \left(\hbar ^4-15 u \hbar ^2+20 u^2\right)\right)\right)+4
   m_1^3 \left(\left(4 u-9 \hbar ^2\right) \left(\left(\hbar ^2-4 u\right)^2 \left(2 \hbar ^4-5 u \hbar ^2+5
   u^2\right)\right.\right.$\\
   &$-\left.\left.12 u \left(\hbar ^4{-}15 u \hbar ^2{+}20 u^2\right) m_3^2\right) \left(4 u{-}\hbar ^2\right)^3{+}12
   \left(4 u{-}9 \hbar ^2\right) m_2^2 \left(4 \left(2 \hbar ^8{-}39 u \hbar ^6{+}371 u^2 \hbar ^4{-}840 u^3 \hbar
   ^2{+}560 u^4\right) m_3^2\right.\right.$\\
   &$-\left.\left.u \left(\hbar ^2-4 u\right)^2 \left(\hbar ^4-15 u \hbar ^2+20 u^2\right)\right)
   \left(4 u-\hbar ^2\right)+4 \left(\hbar ^2-4 u\right)^2 m_2 m_3 \left(-54 \hbar ^{10}+861 u \hbar ^8-6397 u^2
   \hbar ^6\right.\right.$\\
   &$+\left.\left.13100 u^3 \hbar ^4-10480 u^4 \hbar ^2+2880 u^5+4 u \left(27 \hbar ^6-497 u \hbar ^4+1120 u^2 \hbar
   ^2-560 u^3\right) m_3^2\right)\right.$\\
   &$+\left.16 m_2^3 m_3 \left(4 \left(-18 \hbar ^{10}+413 u \hbar ^8-4705 u^2 \hbar
   ^6+15260 u^3 \hbar ^4-18480 u^4 \hbar ^2+6720 u^5\right) m_3^2\right.\right.$\\
   &$-\left.\left.u \left(\hbar ^2-4 u\right)^2 \left(-27 \hbar
   ^6+497 u \hbar ^4-1120 u^2 \hbar ^2+560 u^3\right)\right)\right)\Big]$\\
\end{tabular}
\end{ruledtabular}
\end{table*}

\subsection{Eikonal limit of the a-cycle}

Eq. \eqref{QW} can be mapped to the qSW curve with three flavors \eqref{Q21SW} by introducing the variable
\beq
y=\frac{r-2M}{2M}\,.
\eeq
The dictionary is the following
\bea\label{dict}
q&=&4iM\hbar\sqrt{\omega^2-\frac{\mu^2}{\hbar^2}}\,,\nonumber\\ 
m_1&=&-m_2=2i\hbar M \omega\,, \nonumber\\
\frac{4u}{\hbar^2}&=&4\left(\ell+\frac{1}{2}\right)^2-16M^2\omega^2+8iM\sqrt{\omega^2-\frac{\mu^2}{\hbar^2}}\,,\nonumber\\
m_3&=&\frac{i M \hbar\left(2\omega^2-\frac{\mu^2}{\hbar^2}\right)}{\sqrt{\omega^2-\frac{\mu^2}{\hbar^2}}}\,.
\eea
Replacing   
the dictionary \eqref{dict} in Eq. \eqref{aii}, the eikonal limit can be reproduced via the relations
\beq
\omega=\frac{\mu}{\hbar}\sqrt{1+p_\infty^2},\qquad \ell=\frac{M \mu j}{\hbar}-\frac{1}{2}\,.
\eeq
Expanding for $\hbar,p_\infty\to0$ and we obtain
\bea
-\frac{a-\hbar\left(\ell+\frac{1}{2}\right)}{2M \mu}&=&\frac{\frac{3}{2}+\frac{15p_\infty^2}{8}}{j}+\frac{\frac{35}{8}+\frac{105p_\infty^2}{8}+\frac{1155p_\infty^4}{128}}{j^3}\nonumber\\
&+&\frac{\frac{231}{8}+\frac{9009p_\infty^2}{64}+\frac{27027p_\infty^4}{128}+\frac{51051p_\infty^6}{512}}{j^5}\nonumber\\
&+&O\left( \frac{1}{j^6}\right)\,,
\eea
which are exactly the terms appearing in the $\pi$-sector of the radial action \eqref{Irpi}.
In the next Section we will examine in detail the classical SW curve in the case of three flavours.

\subsubsection{Classical SW curve for $N_f=3$ theory}

Following the notation of \cite{Bianchi:2021mft} we will discuss below how to obtain the prepotential ${\mathcal F}$ from both from the classical curve and
the quantum curve. These results (at least at the classical level) generalize what has been done long ago in Refs.  \cite{Lerche:1996xu,Mironov:2009uv}.

\subsubsection{Obtaining the  prepotential from the classical curve}
The SW curve, Eqs. \eqref{classcurve} and \eqref{PRL0}, in the case of the $N_f=3$ theory is given by
\bea
P_0(x)&=&x^2-u'+q\left(x+\frac{1}{2}-m_1-m_2-m_3\right)\,,\nonumber\\
P_R(x)&{=}&(x{-}m_1)(x{-}m_2),\quad P_L(x){=}x{-}m_3\,.
\eea
After the massless probe limit $\mu=0$, the dictionary \eqref{dict} becomes
\bea
\label{dictQ}
q&=&2i r_s \omega, \quad u=\left(\ell+\frac{1}{2}\right)^2\,,\quad r_s=2M\,,\nonumber\\
m_1&=& -m_2=m_3=\frac{q}{2},\quad u'=u+\frac{q}{2}+\frac{q^2}{4}\,.
\eea
So, using this simplified dictionary, the solution of the classical curve \eqref{classcurve} in the variable $x$ provides
\be
x_\pm=-\frac{q y}{2}\pm \frac{\sqrt{4uy+q^2(1+y)^3}}{2\sqrt{1+y}}\,.
\ee
Consequently we can construct the SW differential
\be
\lambda_0(y)=\frac{x_+(y)-x_-(y)}{4\pi i y}=-i\frac{\sqrt{4uy+q^2(1+y)^3}}{4\pi y\sqrt{1+y}}\,.
\ee
The turning points in this language are the zeros of 
\be
\mathcal{D}=4uy+q^2(1+y)^3=0\,.
\ee
Expanding for small $q$ (or, equivalently, for large $u$), we have
\bea\label{turnn3}
y_1&{=}&\frac{2i\sqrt{u}}{q}{-}\frac{3}{2}{+}\frac{3 i q}{16 \sqrt{u}}{+}\frac{q^2}{8 u}{-}\frac{105 i q^3}{1024 u^{3/2}}\nonumber\\
&{-}&\frac{3 q^4}{32 u^2}{+}\frac{3003 i q^5}{32768 u^{5/2}}{+}\frac{3 q^6}{32 u^3}{-}\frac{415701 i q^7}{4194304
   u^{7/2}}{+}O(q^8)\,,\nonumber\\
y_2&=&-\frac{2i\sqrt{u}}{q}-\frac{3}{2}-\frac{3iq}{16\sqrt{u}}+\frac{q^2}{8 u}+\frac{105 i q^3}{1024 u^{3/2}}\nonumber\\
&{-}&\frac{3 q^4}{32 u^2}{-}\frac{3003 i q^5}{32768 u^{5/2}}{+}\frac{3 q^6}{32 u^3}{+}\frac{415701 i q^7}{4194304
   u^{7/2}}{+}O(q^8)\,,\nonumber\\
y_3&=&-\frac{q^2}{4 u}+\frac{3 q^4}{16 u^2}-\frac{3 q^6}{16 u^3}+O(q^8)\,,
\eea
with $y_1>0>y_2$ and $y_3\in \mathbb{C}$. The integral that corresponds to the scattering radial action is
\be
I(\infty,y_1)=4\pi \int_{\infty}^{y_1}\lambda_0 dy\,.
\ee
Let us introduce the new variable $y=y_1^{(0)}/z$ where $y_1^{(0)}$ is the leading order term in small $q$ expansion of the external turning point $y_1$ in \eqref{turnn3}. This transformation of variable allows us considering only the finite contribution to the integral. 

We can then expand the integrand for small $q$ and integrate order by order from 0 to 1. Finally, we obtain
\bea
&&I(\infty,y_1)=a_{\pi}\pi +a_{\slashed{\pi}}-iq\log(\epsilon)\,,\nonumber\\
a_{\pi}&=&\sqrt{u}+\frac{15 q^2}{64 \sqrt{u}}-\frac{1155 q^4}{16384 u^{3/2}}+\frac{51051 q^6}{1048576 u^{5/2}}\nonumber\\
&-&\frac{47805615 q^8}{1073741824
   u^{7/2}}+\frac{3234846615
   q^{10}}{68719476736 u^{9/2}}+O(q^{12})\,,\nonumber\\
a_{\slashed{\pi}}&{=}&\frac{iq}{2}{-}\frac{i q^3}{3 u}{+}\frac{7 i q^5}{40 u^2}{-}\frac{i q^7}{7 u^3}{+}\frac{143 i q^9}{1008 u^4}{-}\frac{26 i q^{11}}{165 u^5}{+}O(q^{12})\nonumber\\
\eea
where $\epsilon$ is the regulator scale needed to take care of the divergence at $z=0$. The $\pi-$contribution can be exactly resumed as follows
\be
a=a_\pi=\sqrt{u}\,{}_3F_2\Big[\Big\{-\frac{1}{2},\frac{1}{6},\frac{5}{6}\Big\},\Big\{\frac{1}{2},1\Big\},-\frac{27q^2}{u}\Big]\,.
\ee
The previous expression must be inverted considering $q$ small or equivalently large $u$. This task can be accomplished perturbatively
\bea
u&=&a^2-\frac{15 q^2}{32}+\frac{705 q^4}{8192 a^2}-\frac{393 q^6}{16384 a^4}+\frac{6684165 q^8}{536870912 a^6}\nonumber\\
&-&\frac{4091325 q^{10}}{536870912
   a^8}+O(q^{12})\,.
\eea
The prepotential ${\mathcal F}$ is then obtained integrating the Matone relation

\bea\label{Finstnf3}
\mathcal{F}&{=}&{-}\int\frac{u(a)}{q}dq{=}{-}a^2\log(q){+}\frac{15 q^2}{64}{-}\frac{705 q^4}{32768 a^2}\nonumber\\
&{+}&\frac{131 q^6}{32768 a^4}{-}\frac{6684165 q^8}{4294967296
   a^6}{+}\frac{818265 q^{10}}{1073741824
   a^8}{+}O(q^{12})\,.\nonumber\\
\eea

\subsubsection{Obtaining the  prepotential  from the quantum curve}

Using the dictionary \eqref{dictQ} in the continuous fraction relation \eqref{ODEacycl}, we  find (perturbatively) the following expression for $a(u)$
\begin{widetext}
\bea
a(u)&=&\sqrt{u}+\frac{60 u-59}{256 (u-1) \sqrt{u}}q^2+\frac{-73920 u^5+512400
   u^4-1087380 u^3+965435
   u^2-344347 u+27848}{262144
   (u-1)^3 u^{3/2} (u-4) (4 u-1)}q^4\nonumber\\
&+&\frac{q^6}{67108864 (u-9)
   (u-4)^2 (u-1)^5 u^{5/2} (4 u-9)
   (4 u-1)^2}\Big[209104896 u^{11}-4975466496
   u^{10}+46510720256
   u^9\nonumber\\
   &{-}&228254670336
   u^8{+}657185948496
   u^7{-}1171647086456
   u^6{+}1322662323691
   u^5{-}939489611913
   u^4{+}402680065557
   u^3\nonumber\\
   &{-}&94563404971 u^2{+}10214445564
   u{-}532342368\Big]+O(q^8)\,,
\eea
which can be inverted
\bea
u(a)&=&a^2+\frac{59-60 a^2}{128
   \left(a^2-1\right)}q^2+\frac{45120 a^8-306480 a^6+621020
   a^4-503017 a^2+143393}{131072
   \left(a^2-1\right)^3 \left(4
   a^4-17 a^2+4\right)}q^4\nonumber\\
   &+&\frac{q^6}{16777216 (a-3)
   (a-2) (a+2) (a+3) (2 a-3) (2
   a-1) (2 a+1) (2 a+3)
   \left(a^2-1\right)^5}\Big[-6438912 a^{14}\nonumber\\
   &+&120711936
   a^{12}-797796352
   a^{10}+2529456736
   a^8-4329656128 a^6+4113620007
   a^4-2046314138
   a^2\nonumber\\
   &+&416418291\Big]+O(q^8)\,.
\eea
The prepotential ${\mathcal F}$ is then given by
\bea
\mathcal{F}&=&-a^2\log(q)+\frac{\left(60 a^2-59\right)
   q^2}{256 \left(a^2-1\right)}+\frac{\left(-45120 a^8+306480
   a^6-621020 a^4+503017
   a^2-143393\right) q^4}{524288
   (a-2) (a+2) (2 a-1) (2 a+1)
   \left(a^2-1\right)^3}\nonumber\\
   &+&\frac{q^6}{100663296 (a-3) (a-2)
   (a+2) (a+3) (2 a-3) (2 a-1) (2
   a+1) (2 a+3)
   \left(a^2-1\right)^5}\Big[6438912
   a^{14}\nonumber\\
   &-&120711936
   a^{12}+797796352
   a^{10}-2529456736
   a^8+4329656128 a^6-4113620007
   a^4+2046314138
   a^2\nonumber\\
   &-&416418291\Big]+O(q^8)\,.
\eea
In the eikonal limit\footnote{Eikonal limit is done by replacing $a\to a/\hbar$ and $q\to q/\hbar$ and taking the limit $\hbar\to 0$} it becomes
\bea
\mathcal{F}={-}a^2\log(q){+}\frac{15 q^2}{64}{-}\frac{705 q^4}{32768 a^2}{+}\frac{131 q^6}{32768 a^4}{-}\frac{6684165 q^8}{4294967296
   a^6}{+}\frac{818265 q^{10}}{1073741824
   a^8}{-}\frac{7745319363
   q^{12}}{17592186044416 a^{10}}{+}O(q^{12})
\eea
which coincides with the expression given in Eq. \eqref{Finstnf3}.
\end{widetext}

\section{Massive particles  
in the d dimensional Schwarzschild-Tangherlini spacetime}

The   Schwarzschild-Tangherlini metric in $d\ge 4$ dimensions in Schwarzschild-like coordinates writes
\begin{equation}
ds^2
=
-f(r)\,dt^2
+\frac{dr^2}{f(r)}
+r^2 d\Omega_{d-2}^2\,,
\end{equation}
where $d\Omega_{d-2}^2$ is the metric of the unit sphere in $d-2$ dimensions, and
\begin{equation}
f(r)=1-\left(\frac{r_h}{r}\right)^{d-3}\,.
\end{equation}
A widely adopted notation (also used here) is $r_h=M$, which is related to the ADM mass ${\mathcal M}$ of the spacetime
via the relation
\beq
r_h=\left[\frac{16\pi G_d {\mathcal M}}{(d-2)\Omega_{d-2}}\right]^{\frac{1}{d-3}}\,,
\eeq
where  $G_d$ denotes the $d$-dimensional gravitational constant and $\Omega_{d-2}$ is the volume of the unit sphere in dimensions $d-2$ 
\beq
\Omega_{d-2}=\frac{2\pi^{\frac{d-1}{2}}}{\Gamma \left( \frac{d-1}{2}\right)}\,.
\eeq

Let us limit to massive geodesics confined to the equatorial hyperplane $\theta=\pi/2$, namely exploiting the spherical symmetry of the spacetime 
one can always consider the limit 
\beq
d\Omega_{d-2}^2\to d\phi^2
\eeq
which reduces the metric to the form
\beq
ds_{\rm equat}^2=-f(r)dt^2+\frac{dr^2}{f(r)}+r^2d\phi^2\,,  
\eeq
also assuming that the (constant Killing related) conjugate momenta to all the angular variables vanish except for $(\phi,P_\phi)$, for simplicity and aiming at 
staying as close as possible to the 4 dimensional Schwarzschild spacetime.

The Hamilton mass-shell condition is
\beq
g^{\mu\nu}P_{\mu}P_{\nu}=-\mu^2\,.
\eeq
where the conserved momenta are
\beq
P_t=-E,\quad P_\phi=J\,.
\eeq
Let us introduce
\beq
E=\mu\sqrt{1+2\bar{E}}\,,\quad J=\mu M j
\eeq
The radial effective potential reads
\bea
P_r^2&=&Q_r(r,j)\nn\\
Q_r(r,j)&{=}&\frac{r^{d{-}2}}{(M^dr^3{-}M^3r^d)^2}\Big[M^{d{+}3}r^5{+}2\bar{E}M^6r^{2{+}d}\nn\\
&{+}&j^2M^5(M^dr^3{-}M^3r^d)\Big]\,.
\eea
Introducing
\beq
\bar{E}=\frac{p_\infty^2}{2},\quad j=\frac{p_\infty b}{M},\quad b=\hat{b}M
\eeq
we can identify two \lq\lq turning ponts'': an internal turning point and an external one. The large-$b$ expansion limit of the internal turning point reads
\bea
r_-&=&M\Big[1+\frac{M^2}{b^2}\left(\frac{1}{(d-3) p_{\infty
   }^2}+\frac{1}{d-3}\right)\nn\\
   &+&\frac{M^4}{b^4}\left(\frac{5 }{(d-3)^2 p_{\infty
   }^2}-\frac{(d-8) }{2
   (d-3)^2 p_{\infty
   }^4}+\frac{(d+2) }{2
   (d-3)^2}\right)\nn\\
   &+&\frac{M^6}{b^6}\left(-\frac{7 (d-10)}{2 (d-3)^3
   p_{\infty }^4}+\frac{(2 d-13)
   (d-10)}{6 (d-3)^3 p_{\infty
   }^6}\right.\nn\\
   &{+}&\left.\frac{7 (d{+}4)}{2 (d{-}3)^3
   p_{\infty }^2}{+}\frac{(d{+}4) (2
   d{+}1)}{6 (d{-}3)^3}\right)\Big]{+}O\Big[\left(\frac{M}{b}\right)^8\Big]\,,\nn\\
\eea
suggesting, for example, $b\to b \sqrt{d-3}$ to have more compact expressions (recalling that we are interested only in $d\ge 4$),
while for the external turning point one finds
\bea\label{rpd}
r_+&=&b\Big[1+\left(-\frac{1}{2 p_{\infty
   }^2}-\frac{1}{2}\right)\left(\frac{M}{b}\right)^{d-3}\nn\\
   &+&\left(\frac{7-2 d}{4 p_{\infty
   }^2}+\frac{9-2 d}{8 p_{\infty
   }^4}+\frac{1}{8} (5-2 d)\right)\left(\frac{M}{b}\right)^{2(d-3)}\nn\\
   &+&\left(\frac{-\frac{9 d^2}{16}+\frac{27
   d}{8}-5}{p_{\infty
   }^2}\right.\nn\\
   &-&\left.\frac{3 \left(3 d^2-22
   d+40\right)}{16 p_{\infty
   }^4}+\frac{-3 d^2+26 d-56}{16
   p_{\infty }^6}-\frac{3
   d^2}{16}\right.\nn\\
   &+&\left.\frac{7 d}{8}-1\right)\left(\frac{M}{b}\right)^{3(d-3)}\nn\\
   &+&\left(\frac{-64 d^3+528 d^2-1436
   d+1287}{96 p_{\infty
   }^2}\right.\nn\\
   &+&\left.\frac{-64 d^3+624
   d^2-2012 d+2145}{64 p_{\infty
   }^4}\right.\nn\\
   &+&\left.\frac{-64 d^3+720
   d^2-2684 d+3315}{96 p_{\infty
   }^6}\right.\nn\\
   &+&\left.\frac{-64 d^3+816
   d^2-3452 d+4845}{384
   p_{\infty }^8}\right.\nn\\
   &{+}&\left.\frac{1}{384}
   \left({-}64 d^3{+}432 d^2{-}956
   d{+}693\right)\right)\left(\frac{M}{b}\right)^{4(d{-}3)}\Big]\nn\\
   &+&O\Big[\left(\frac{M}{b}\right)^{5(d-3)}\Big]\,.
\eea
Passing to the new variable
\beq
r=\frac{r_+^{(0)}}{u}
\eeq
where $r_+^{(0)}$ represents the leading order of \eqref{rpd}, which is enough to select the finite part of the radial action integral~\cite{Damour:1988mr}.

The radial action integral becomes
\beq
I_r^{(d)}=\hat{b}Mp_\infty\int_0^1 du\frac{\sqrt{1-u^2}\sqrt{1+\frac{u^{d+3}(1+p_\infty^2u^2)}{\hat{b}^{d-3}p_\infty^2(1-u^2)}}}{u^2\left(1-\frac{u^{d-3}}{\hat{b}^{d-3}}\right)}\,.
\eeq
For large $\hat{b}$ the previous integral can be expanded integrated and then (non-trivially) resummed in terms of Fox-Wright functions 
\begin{widetext}
     \beq
I_r^{(d)}=
\frac{1}{2}p_\infty \hat{b}M\sqrt{\pi}\sum_{w=0}^\infty \frac{(\frac{-1}{p_\infty^2})^w}{w!}
\,{}_4\Psi_4\!\left[
\begin{array}{c}
\left(1,1\right),
\left(\frac{1}{2},\frac{1-d}{2}\right),
\left(\frac{1}{2}+\frac{d-1}{2}\right),
\left(-1,d-3\right)\\[2mm]
\left(0,d-3\right),
\left(1-w,1\right),
\left(1-w,\frac{d-3}{2}\right),
\left(\frac{1}{2}+w,\frac{1-d}{2}\right)
\end{array}
;\,
\hat{b}^{\,3-d}
\right]\,.
\eeq
In expanded form
\bea
I_r^{(d)}&=&-\frac{1}{2} \pi  \hat{b} M
   p_{\infty }+\hat{b}^{4-d}
   \left(\frac{\sqrt{\pi } M
   \Gamma
   \left(\frac{d}{2}\right)
   p_{\infty }}{2 (d-4) \Gamma
   \left(\frac{d-1}{2}\right)}+\frac{\pi ^{3/2} M \csc
   \left(\frac{\pi 
   d}{2}\right)}{4 \Gamma
   \left(3-\frac{d}{2}\right)
   \Gamma
   \left(\frac{d-3}{2}\right)
   p_{\infty }}\right)+\hat{b}^{7-2 d}
   \left(\frac{\sqrt{\pi } M
   \Gamma
   \left(d-\frac{1}{2}\right)
   p_{\infty }}{4 (2 d-7) \Gamma
   (d-2)}\right.\nn\\
   &+&\left.\frac{\pi ^{3/2}
   (-1)^d M}{2 (2 d-7) \Gamma
   \left(\frac{5}{2}-d\right)
   \Gamma (d-3) p_{\infty
   }}+\frac{\pi ^{3/2} (-1)^d
   M}{8 \Gamma
   \left(\frac{9}{2}-d\right)
   \Gamma (d-4) p_{\infty
   }^3}\right)+\hat{b}^{10-3 d}
   \left(\frac{\sqrt{\pi } M
   \Gamma \left(\frac{3
   d}{2}-1\right) p_{\infty
   }}{12 (3 d-10) \Gamma
   \left(\frac{3
   d}{2}-\frac{7}{2}\right)}\right.\nn\\
   &+&\left.\frac{\pi ^{3/2} M \csc
   \left(\frac{3 \pi 
   d}{2}\right)}{4 (3 d-10)
   \Gamma \left(3-\frac{3
   d}{2}\right) \Gamma
   \left(\frac{3
   (d-3)}{2}\right) p_{\infty
   }}-\frac{\pi ^{3/2} M \csc
   \left(\frac{3 \pi 
   d}{2}\right)}{4 (3 d-10)
   \Gamma \left(4-\frac{3
   d}{2}\right) \Gamma
   \left(\frac{3
   d}{2}-\frac{11}{2}\right)
   p_{\infty }^3}-\frac{\pi
   ^{3/2} M \csc \left(\frac{3
   \pi  d}{2}\right)}{24 \Gamma
   \left(6-\frac{3 d}{2}\right)
   \Gamma \left(\frac{3
   d}{2}-\frac{13}{2}\right)
   p_{\infty }^5}\right)\nn\\
   &+&\hat{b}^{13-4 d}
   \left(\frac{\sqrt{\pi } M
   \Gamma \left(2
   d-\frac{3}{2}\right)
   p_{\infty }}{48 (4 d-13)
   \Gamma (2 d-5)}-\frac{\pi
   ^{3/2} M}{12 (4 d-13) \Gamma
   \left(\frac{7}{2}-2 d\right)
   \Gamma (2 d-6) p_{\infty
   }}+\frac{\pi ^{3/2} M}{8 (4
   d-13) \Gamma
   \left(\frac{9}{2}-2 d\right)
   \Gamma (2 d-7) p_{\infty
   }^3}\right.\nn\\
   &-&\left.\frac{\pi ^{3/2} M}{12
   (4 d-13) \Gamma
   \left(\frac{11}{2}-2 d\right)
   \Gamma (2 d-8) p_{\infty
   }^5}-\frac{\pi ^{3/2} M}{96
   \Gamma \left(\frac{15}{2}-2
   d\right) \Gamma (2 d-9)
   p_{\infty }^7}\right)+O\left(\hat{b}^{16-5d}\right)\,,
\eea
such that the scattering angle which follows by (the negative of) differentiation with respect to $j$  (it its resummed form) is given by
    \beq
\frac{ \chi^{(d)}+\pi}{2}=
\frac{\sqrt{\pi}}{2}\sum_{w=0}^\infty \frac{(\frac{-1}{p_\infty^2})^w}{w!}
\,{}_3\Psi_3\!\left[
\begin{array}{c}
\left(1,1\right),
\left(\frac{1}{2},\frac{1-d}{2}\right),
\left(\frac{1}{2},\frac{d-1}{2}\right)\\[2mm]
\left(1-w,1\right),
\left(1-w,\frac{d-3}{2}\right),
\left(\frac{1}{2}+w,\frac{1-d}{2}\right)
\end{array}
;\,
\hat{b}^{\,3-d}
\right]\,.
\eeq
Note that the various arguments of the Fox-Wright representation of the radial action and of the scattering coincide and, practically, the $j$ differentiation has lowered simply the order of the Fox-Wright  function.

In the expanded form we have then
\bea
\frac{ \chi^{(d)}+\pi}{2}=\frac{\pi}{2}+C_1 \frac{1}{(\hat b)^{(d-3)}}+C_2 \frac{1}{(\hat b)^{2(d-3)}}+C_3 \frac{1}{(\hat b)^{3(d-3)}}+C_4 \frac{1}{(\hat b)^{4(d-3)}}+C_5 \frac{1}{(\hat b)^{5(d-3)}}+\ldots
\eea
where
\bea
C_1 &=& \frac{\sqrt{\pi } \Gamma
   \left(\frac{d}{2}\right)}{2
   \Gamma
   \left(\frac{d-1}{2}\right)}-\frac{\pi ^{3/2} \csc
   \left(\frac{\pi
   d}{2}\right)}{2 p_\infty^2
   \Gamma
   \left(2-\frac{d}{2}\right)
   \Gamma
   \left(\frac{d-3}{2}\right)}\,,\nonumber\\
C_2 &=& -\frac{\pi ^{3/2} \sec (\pi
   d)}{4 p_\infty^4 \Gamma
   \left(\frac{7}{2}-d\right)
   \Gamma (d-4)}+\frac{\pi
   ^{3/2} \sec (\pi  d)}{2
   p_\infty^2 \Gamma
   \left(\frac{5}{2}-d\right)
   \Gamma (d-3)}+\frac{\sqrt{\pi
   } \Gamma
   \left(d-\frac{1}{2}\right)}{4
   \Gamma (d-2)}\,,\nonumber\\ 
C_3 &=& \frac{\pi ^{3/2} \csc
   \left(\frac{3 \pi
   d}{2}\right)}{12
   p_\infty^6 \Gamma
   \left(5-\frac{3 d}{2}\right)
   \Gamma \left(\frac{3
   d}{2}-\frac{13}{2}\right)}-\frac{\pi ^{3/2} \csc
   \left(\frac{3 \pi
   d}{2}\right)}{4 p_\infty^4
   \Gamma \left(4-\frac{3
   d}{2}\right) \Gamma
   \left(\frac{3
   d}{2}-\frac{11}{2}\right)}+\frac{\pi ^{3/2} \csc
   \left(\frac{3 \pi
   d}{2}\right)}{4 p_\infty^2
   \Gamma \left(3-\frac{3
   d}{2}\right) \Gamma
   \left(\frac{3
   (d-3)}{2}\right)}+\frac{\sqrt
   {\pi } \Gamma \left(\frac{3
   d}{2}-1\right)}{12 \Gamma
   \left(\frac{3
   d}{2}-\frac{7}{2}\right)}\,,\nonumber\\
C_4 &=& \frac{\pi ^{3/2} \sec (2 \pi
   d)}{48 p_\infty^8 \Gamma
   \left(\frac{13}{2}-2 d\right)
   \Gamma (2 d-9)}-\frac{\pi
   ^{3/2} \sec (2 \pi  d)}{12
   p_\infty^6 \Gamma
   \left(\frac{11}{2}-2 d\right)
   \Gamma (2 d-8)}\nonumber\\
&+&\frac{\pi
   ^{3/2} \sec (2 \pi  d)}{8
   p_\infty^4 \Gamma
   \left(\frac{9}{2}-2 d\right)
   \Gamma (2 d-7)}-\frac{\pi
   ^{3/2} \sec (2 \pi  d)}{12
   p_\infty^2 \Gamma
   \left(\frac{7}{2}-2 d\right)
   \Gamma (2
   d-6)}+\frac{\sqrt{\pi }
   \Gamma \left(2
   d-\frac{3}{2}\right)}{48
   \Gamma (2 d-5)}\,,\nonumber\\
C_5 &=& -\frac{\pi ^{3/2} \csc
   \left(\frac{5 \pi
   d}{2}\right)}{240
   p_\infty^{10} \Gamma
   \left(8-\frac{5 d}{2}\right)
   \Gamma \left(\frac{5
   d}{2}-\frac{23}{2}\right)}+\frac{\pi ^{3/2} \csc
   \left(\frac{5 \pi
   d}{2}\right)}{48
   p_\infty^8 \Gamma
   \left(7-\frac{5 d}{2}\right)
   \Gamma \left(\frac{5
   d}{2}-\frac{21}{2}\right)}-\frac{\pi ^{3/2} \csc
   \left(\frac{5 \pi
   d}{2}\right)}{24
   p_\infty^6 \Gamma
   \left(6-\frac{5 d}{2}\right)
   \Gamma \left(\frac{5
   d}{2}-\frac{19}{2}\right)}\nonumber\\
&+& \frac{\pi ^{3/2} \csc
   \left(\frac{5 \pi
   d}{2}\right)}{24
   p_\infty^4 \Gamma
   \left(5-\frac{5 d}{2}\right)
   \Gamma \left(\frac{5
   d}{2}-\frac{17}{2}\right)}-\frac{\pi ^{3/2} \csc
   \left(\frac{5 \pi
   d}{2}\right)}{48
   p_\infty^2 \Gamma
   \left(4-\frac{5 d}{2}\right)
   \Gamma \left(\frac{5
   (d-3)}{2}\right)}+\frac{\sqrt
   {\pi } \Gamma \left(\frac{5
   d}{2}-2\right)}{240 \Gamma
   \left(\frac{5
   (d-3)}{2}+1\right)}
\,.
\eea

\end{widetext}

Having identified the Fox-Wright functions here has been possible not only because of technical virtuous work, but especially because 
the occurrence of Fox-Wright functions in the radial action was already noticed when computing it for photons in a recent work \cite{Bini:2026foz}, which was driving us in our research here.
With the present confirmation, the presence of the Fox-Wright functions  seems naturally associated with the wider picture offered when increasing the spacetime dimensions.

A final remark concerns the fact that specific values of the dimension $d$ may simplify much the above expressions or even generate some divergent behaviour. We would like to remind the reader that the ultimate value of the dimension $d$ we are actually interested in is $d=4+\epsilon$, with a successive series in small $\epsilon$, in order to  proceed with dimensional regularization of observables naturally living in $d=4$. 

In this specific case we find $I_r^\epsilon\equiv I_r^{d=4+\epsilon}$ such that
\bea
I_r^\epsilon=\frac{I_r^{-1}}{\epsilon}+I_r^0+ I_r^1\epsilon +O(\epsilon^2)\,,
\eea
where 
\beq
I_r^{-1}=I^{\slashed{\pi}}_0 \,,
\eeq
as form Eq. \eqref{I_non_pi_defs},
the order in zero in $\epsilon$ coincides with the 4d Schwarzschild \eqref{Irpi} and \eqref{Irpi2} and
\bea
I_r^{1}&=& I_r^{1\,{\ln^0}}+I_r^{1  \ln^1{}}\ln \left(\frac{j}{4p_\infty}\right) \nn\\
&+& I_r^{1  \ln^2{}}\ln^2 \left(\frac{j}{4p_\infty}\right) \,, 
\eea
with
\begin{widetext}
\bea
I_r^{1\,{\ln^0}}
&=& \frac{\pi  \left(\frac{11017 p_{\infty }^4}{384}+\frac{601 p_{\infty
   }^2}{12}+\frac{5}{3 p_{\infty }^2}-\frac{1}{6 p_{\infty
   }^4}+\frac{47}{2}\right)}{j^3}-4\log (2) \left(\frac{I_1^\pi}{j} +\frac{2I_3^{\pi}}{j^3}\right)
\nonumber\\
&+&\frac{-\frac{76 p_{\infty }^3}{3}-21
   p_{\infty }+\frac{3}{2 p_{\infty }}+\frac{1}{2 p_{\infty
   }^3}}{j^2}+\frac{\pi  \left(\frac{p_{\infty }^2}{8}+\frac{1}{2 p_{\infty
   }^2}+1\right)}{j}-\frac{\pi ^2}{24} I_0^{\slashed \pi} +p_{\infty }\nonumber\\ 
I_r^{1  \ln^1}&=& -\frac{\pi I_3^\pi }{2j^3}-\frac{3 I_2^{\slashed \pi}}{4j^2}-\frac{\pi I_1^\pi}{j}+p_{\infty }
\nonumber \\
I_r^{1  \ln^2}&=& 
\frac12  I_0^{\slashed \pi}\,,
\eea
\end{widetext}
having replaced $M\to 2M$ to make a direct comparison with the standard Schwarzschild solution.

We will postpone to future works further investigations on the limiting behaviours of the Fox-Wright functions when $d$ collapses to the value $d=4+\epsilon$, as well as related considerations on this very interesting topic. 

\section{D3 branes spacetimes}
 
Let us focus our attention on the propagation of massless probes in the D3-branes geometry. This choice is motivated purely by their simplicity and by the resulting technical advantages. Massless waves propagating in D3 branes are always described by a particular confluent form of the Heun equation, more precisely, by the so-called Doubly Reduced Doubly Confluent Heun Equation  \cite{Bianchi:2022wku,Bonelli:2022ten}, well known in the mathematical literature as the Mathieu equation \cite{Bianchi:2021xpr,Gregori:2022xks,Fioravanti:2021dce}.

The idea is the following: since the \lq\lq scattering" radial action, i.e. the radial action integral evaluated over the external classically allowed region, is related to the SW $a-$cycle, the \lq\lq instantonic" radial action integral, namely the one evaluated in the classically forbidden region, must be related to the instantonic contribution to the quantum version of the SW $a_D$ \cite{Fioravanti:2019vxi}.

After establishing this relation, the resummation properties in the eikonal limit derived from the geodesics can be transferred to the wave side.

\subsection{Geodesic motion in D3 branes geometry}

The D3 branes metric in 10 dimensions reads
\beq
ds^2=H(r)^{-\frac{1}{2}}(-dt^2+d\mathbf{x}^2)+H(r)^{\frac{1}{2}}(dr^2+r^2d\Omega_5^2)\,,
\eeq
where $\mathbf{x}$ are the longitudinal conserved coordinates, $H(r)=1+\frac{L^4}{r^4}$ is the relevant harmonic function with $L$ the D3-brane charge, and $d\Omega_5^2$ denotes the metric of the $S^5$ sphere. In the equatorial plane, and dropping all the longitudinal directions contributions, we are left with
\beq
ds^2=-\left(1+\frac{L^4}{r^4}\right)^{-\frac{1}{2}}dt^2+\left(1+\frac{L^4}{r^4}\right)^{\frac{1}{2}}(dr^2+r^2d\phi^2)\,.
\eeq
These geometries are known to be \lq \lq small" BHs, since their event horizon radius is located at $r=0$.

The mass shell condition for geodesic motion reads
\be
\mathcal{H}=g^{\mu\nu}P_\mu P_\nu=0\,,
\ee
and can be solved for the radial momentum as follows
\be
p_r=\frac{P_r}{E}=\frac{\sqrt{r^4-b^2r^2+L^4}}{r^2}\,.
\ee
From the condition $p_r=0$, we identify the \lq\lq turning points"
\be
r_\pm=\sqrt{\frac{b^2\pm\sqrt{b^4-4L^4}}{2}}\,,
\ee
so that the condition $b>\sqrt{2}L$ implies $r_+>r_->0$. The photon sphere is located at
\be
p_r=0=\frac{dp_r}{dr}\,,
\ee
which is solved by $r_c=L$ and $b_c=\sqrt{2}L$.

In the classically allowed region, the radial action of a massless particle moving between the two generic radii $r_1 > r_2 > 0$ is defined as
\be
I_r(r_1,r_2)=\int_{r_2}^{r_1} p_r \, dr\,.
\ee
As a non trivial result it holds
\bea
I_r(r_+,\infty)=I_r(0,r_-)=I_r^{\rm scatt}\,,
\eea
as a consequence of the (not yet sufficiently understood) Couch-Torrence (CT) symmetry \cite{CT1984}. CT symmetry was originally discovered in the context of extremal Reissner-Nordstr\"om  geometry. In particular, the extremal Reissner-Nordstr\"om metric is invariant, up to an overall \lq\lq Weyl factor," under an inversion symmetry that exchanges radial infinity with the horizon while keeping fixed the focal point of the photon sphere. Even though this symmetry is peculiar to BPS geometries, its origin remains unknown. 

Differently from the \lq\lq boundary-to-bound" correspondence in binary processes between the periastron advance $\Delta\phi(J,E)$ and the scattering angle $\chi(J,E)$ \cite{Kalin:2019rwq,Kalin:2019inp,Cho:2021arx,Bini:2020nsb}
\be
\Delta\phi(J,E)=\chi(J,E)+\chi(-J,E)\quad,\quad E<0
\ee
which requires analytic continuation in the binding energy $E$ and angular momentum $J$, CT symmetry maps the two geodesics allowed for $b>b_c$ ($b_c=\sqrt{2}L$ for D3-branes) without requiring any analytic continuation. However, CT symmetry seems to hold only in a restricted class of cases, such as D3-branes, D1–D5, and D3D3D3D3 systems \cite{Bianchi:2021yqs}\footnote{In the latter case, precise conditions on the brane charges must be satisfied in order for CT symmetry to hold}.

CT symmetry in rotating systems has a different meaning and different consequences: it is no longer a Weyl symmetry of the metric under conformal inversion. Instead, the symmetry is much weaker and emerges as a symmetry of the radial wave equation under the same transformation as in the spherically symmetric case. This situation has been fully analyzed for extremal Kerr–Newman BHs, as well as for rotating generalizations of D3-branes, in \cite{Cvetic:2020kwf,Bianchi:2022wku}. Since CT symmetry acts on the horizon, it is reasonable that it is characteristic of BHs and cannot hold for smooth, horizonless geometries such as fuzzballs.

For the D3 branes case CT symmetry reads
\be
r\to r'=\frac{L^2}{r}\,.
\ee
As a consequence of CT, notice that
\be
r_\pm=\frac{L^2}{r_\mp}\,,
\ee
and the integrand of the radial action transforms as
\be
p_r dr\underset{r'=\frac{L^2}{r}}{\longrightarrow}-p_r' dr'\,.
\ee
So that, after regularization, we have 
\bea
I_r^{\rm scatt}&=&2r_+E\left(\frac{r_-^2}{r_+^2}\right)-\frac{r_+^2-r_-^2}{r_+}K\left(\frac{r_-^2}{r_+^2}\right)\nonumber\\
&=&\frac{b\pi}{2}{}_2F_1\left(-\frac{1}{4},\frac{1}{4},1,\frac{4L^4}{b^4}\right)\nonumber\\
&=&\frac{b\pi}{2}\left[1-\frac{L^4}{4 b^4}-\frac{15 L^8}{64 b^8}-\frac{105 L^{12}}{256 b^{12}}\right.\nonumber\\
&-&\left.\frac{15015 L^{16}}{16384
   b^{16}}{-}\frac{153153 L^{20}}{65536
   b^{20}}{+}O\left(\frac{1}{b^{24}}\right)\right]\,,\nonumber\\
\eea
where the complete elliptic integrals are defined
\bea
E(k)&=&\int_0^{\pi/2}(1-k \sin^2\theta)^{1/2}d\theta\,,\nonumber\\
K(k)&=&\int_0^{\pi/2}(1-k \sin^2\theta)^{-1/2}d\theta\,,
\eea
which can be related to
\bea
E(k)&=&\frac{\pi}{2}{}_2F_1\left(\frac{1}{2},-\frac{1}{2},1,k\right)\,,\nonumber\\
K(k)&=&\frac{\pi}{2}{}_2F_1\left(\frac{1}{2},\frac{1}{2},1,k\right)\,.
\eea

The \lq \lq instantonic" radial action can be exactly integrated as
\bea
I_r^{\rm inst}&=&I_r(r_+,r_-)=\int_{r_-}^{r_+}p_r \, dr\nonumber\\
&=&i\Bigg[{-}2r_+ E\left(1{-}\frac{r_-^2}{r_+^2}\right){+}\frac{r_+^2{+}r_-^2}{r_+}K\left(1{-}\frac{r_-^2}{r_+^2}\right)\Bigg]\nonumber\\
\eea
which is expanded in PM sense (large $b$) as follows
\begin{widetext}
    \bea\label{adgeo1}
I_r^{\rm inst}&=&-2ib\left(1-\frac{L^4}{8 b^4}-\frac{47 L^8}{256 b^8}-\frac{1097 L^{12}}{3072 b^{12}}-\frac{329177 L^{16}}{393216
   b^{16}}-\frac{1150433 L^{20}}{524288
   b^{20}}+O\left(\frac{1}{b^{24}}\right)\right)+\nonumber\\
   &+&2i b\log\left(\frac{2b}{L}\right)\left( 1-\frac{L^4}{4 b^4}-\frac{15 L^8}{64 b^8}-\frac{105 L^{12}}{256 b^{12}}-\frac{15015 L^{16}}{16384
   b^{16}}-\frac{153153 L^{20}}{65536
   b^{20}}+O\left(\frac{1}{b^{24}}\right)\right)\,.
    \eea
    Notice that the coefficient of the log-term in $I_r^{\rm inst}$ is proportional to $I_r^{\rm scatt}$, so that this sector of the instantonic radial action is already resummed.
\end{widetext}
\eqref{adgeo1} can be resummed as follows \cite{Lerche:1996xu}
\beq
I_r^{\rm inst}=-\frac{i\pi L}{4}\left(1-\frac{b^4}{4L^4}\right){}_2F_1\left(\frac{3}{4},\frac{3}{4},2,1-\frac{b^4}{4L^4}\right)\,.
\eeq

\subsection{qSW for D3-branes}

The doubly-reduced doubly-confluent Heun equation is simplified since, from Eq. \eqref{PRL0},  we have
\be\label{parcurve}
P_L(x)=P_R(x)=1\quad,\quad P_0(x)=x^2-u\,.
\ee
The qSW curve with zero flavors is 
\be
Q_{0}(y)=\frac{4y^2q+y(1-4u)+4}{4y^3}\,.
\ee
The continuous fraction \eqref{ODEacycl} is 
solved perturbatively in $a$ as follows
\begin{widetext}
\bea
a&=&\sqrt{u}{+}\frac{1}{\sqrt{u} \left(\hbar^2{-}4 u\right)}q{+}\frac{{-}60 u^2{+}35 u \hbar^2{-}2 \hbar^4}{4 u^{3/2} \left(u{-}\hbar^2\right) \left(4
   u{-}\hbar^2\right)^3}q^2{+}\frac{{-}6720 u^5{+}18480 u^4 \hbar^2{-}15260 u^3 \hbar^4{+}4705 u^2 \hbar^6{-}413 u
   \hbar^8{+}18 \hbar^{10}}{4 u^{5/2} \left(4 u{-}9 \hbar^2\right) \left(4 u{-}\hbar^2\right)^5 \left(\hbar ^2{-}u\right)^2}q^3\nonumber\\
   &+&\frac{q^4}{64 u^{7/2} \left(u{-}4 \hbar^2\right) \left(u{-}\hbar^2\right)^3
   \left(4 u{-}\hbar^2\right)^7 \left(9 \hbar^2{-}4 u\right)^2}\Bigg[5 \left({-}3075072 u^9{+}25369344 u^8 \hbar^2{-}79567488 u^7 \hbar^4{+}123135584
   u^6 \hbar^6\right.\nonumber\\
   &{-}&\left.101477948 u^5 \hbar^8{+}45396975 u^4 \hbar^{10}{-}10473935 u^3
   \hbar^{12}{+}1135160 u^2 \hbar^{14}{-}80712 u \hbar^{16}{+}2592 \hbar^{18}\right)\Bigg]+O(q^5)
\eea
One can invert the previous equation perturbatively in order to reconstruct the Coulomb branch modulus parameter $u$ as a function of $a$.
\bea
u&=&a^2+\frac{2}{4 a^2-\hbar ^2}q+\frac{20 a^2+7 \hbar ^2}{2 \left(a^2-\hbar ^2\right) \left(4 a^2-\hbar
   ^2\right)^3}q^2+\frac{4 \left(144 a^4+232 a^2 \hbar ^2+29 \hbar ^4\right)}{\left(4 a^2-\hbar
   ^2\right)^5 \left(4 a^4-13 a^2 \hbar ^2+9 \hbar ^4\right)}q^3\nonumber\\
   &+&\frac{376064 a^{10}+585216 a^8 \hbar ^2-2245664 a^6 \hbar ^4+256912 a^4 \hbar
   ^6+827565 a^2 \hbar ^8+68687 \hbar ^{10}}{32 \left(a^2-\hbar ^2\right)^3
   \left(4 a^2-\hbar ^2\right)^7 \left(4 a^4-25 a^2 \hbar ^2+36 \hbar ^4\right)}q^4+O(q^5)\,.
\eea
\end{widetext}
Now one can use the Matone relation \cite{Matone:1995rx,Flume:2004rp} to find the prepotential $\mathcal{F}$
\be
u=-q\frac{\partial \mathcal{F}}{\partial q}\,.
\ee
The dual cycle $a_D$ is then defined as
\be
a_D=\frac{\partial \mathcal{F}}{\partial a}\,.
\ee
The quantum Matone relation provides the three level and the instantonic contribution to the prepotential. 
\begin{widetext}
The prepotential expanded for small $\hbar$
\bea
\mathcal{F}&=&-a^2\log(q)-\frac{q}{2 a^2}-\frac{5 q^2}{64 a^6}-\frac{3 q^3}{64 a^{10}}-\frac{1469 q^4}{32768 a^{14}}-\frac{4471 q^5}{81920 a^{18}}-\frac{40397 q^6}{524288 a^{22}}+O(q^7)\nonumber\\
&+&\hbar^2\left(-\frac{q}{8 a^4}-\frac{21 q^2}{128 a^8}-\frac{55 q^3}{192 a^{12}}-\frac{18445 q^4}{32768 a^{16}}-\frac{193401 q^5}{163840 a^{20}}-\frac{4057361 q^6}{1572864
   a^{24}}+O(q^6)\right)+O(\hbar^4)\,,
\eea
and
\bea\label{faNS}
\partial_a\mathcal{F}&=&-2a\log(q)+\frac{q}{a^3}+\frac{15 q^2}{32 a^7}+\frac{15q^3}{32a^{11}}+\frac{10283 q^4}{16384 a^{15}}+\frac{40239 q^5}{40960 a^{19}}+\frac{444367 q^6}{262144 a^{23}}+O(q^6)\nonumber\\
&+&\hbar^2\left(\frac{q}{2 a^5}+\frac{21 q^2}{16 a^9}+\frac{55 q^3}{16 a^{13}}+\frac{18445 q^4}{2048 a^{17}}+\frac{193401 q^5}{8192 a^{21}}+\frac{4057361 q^6}{65536 a^{25}}+O(q^6)\right)+O(\hbar^4)\,.
\eea
\end{widetext}

\subsection{Classical SW curve}

From Eq. \eqref{parcurve}, the classical curve reads
\be
qy^2P_L(x)+yP_0(x)+P_R(x)=0\,,
\ee
and can be solved in the  $x$  variable as follows
\be
x_\pm(y)=\pm\sqrt{\frac{uy-1-qy^2}{y}}\,.
\ee
The SW differential is
\be
\lambda_0(x)=\frac{x_+(y)-x_-(y)}{4\pi i y}\,,
\ee
and the turning points are given by
\be
y_\pm=\frac{u\pm\sqrt{u^2-4q}}{2q}\,.
\ee
The fundamental cycle is defined as
\be
a=\int_\infty^{y_1}\lambda_0(y) dy\,,
\ee
which, in terms of the new variable $y=y_+/z$, becomes
\be
a=-\frac{\sqrt{q y_-}}{2\pi}\int_0^1\frac{\sqrt{1-v}\sqrt{\frac{y_+}{y_-}-v}}{v^{3/2}}dv\,.
\ee
The finite part of the previous integral is then given by
\be\label{invert}
a(u)=\sqrt{u}{}_2F_1\left(-\frac{1}{4},\frac{1}{4},1,\frac{4q}{u^2}\right)\,,
\ee
and can be expanded as follows
\bea
a(u)&=&\sqrt{u}-\frac{q}{4 u^{3/2}}-\frac{15 q^2}{64 u^{7/2}}-\frac{105 q^3}{256 u^{11/2}}\nonumber\\
&-&\frac{15015 q^4}{16384 u^{15/2}}-\frac{153153 q^5}{65536 u^{19/2}}-\frac{6789783 q^6}{1048576 u^{23/2}}\nonumber\\
&-&\frac{79676025 q^7}{4194304 u^{27/2}}-\frac{62386327575 q^8}{1073741824 u^{31/2}}+O(q^9)\,.\nonumber\\
\eea
Inverting \eqref{invert} formally at any order  is still challenging. Proceeding order by order in $q$, instead, the previous expression can be inverted as follows
\bea
u(a)&=&a^2+\frac{q}{2 a^2}+\frac{5 q^2}{32 a^6}+\frac{9 q^3}{64 a^{10}}+\frac{1469 q^4}{8192 a^{14}}\nonumber\\
&+&\frac{4471 q^5}{16384 a^{18}}+\frac{121191 q^6}{262144 a^{22}}+\frac{441325 q^7}{524288 a^{26}}\nonumber\\
&+&\frac{866589165 q^8}{536870912 a^{30}}+O(q^9)\,.
\eea
Using the Matone relation
\beq
-2\pi i a_D=-\partial_a\int\frac{u(a)}{q}dq=\partial_a\mathcal{F}\,,
\eeq
we find
\bea
\partial_a\mathcal{F}&=&-2a\log(q)+\frac{q}{a^3}+\frac{15 q^2}{32 a^7}+\frac{15 q^3}{32 a^{11}}\nonumber\\
&+&\frac{10283 q^4}{16384 a^{15}}+\frac{40239 q^5}{40960 a^{19}}+\frac{444367 q^6}{262144 a^{23}}\nonumber\\
&+&\frac{5737225 q^7}{1835008 a^{27}}+\frac{12998837475 q^8}{2147483648 a^{31}}+O(q^9)\nonumber\\
\eea
which is exactly the order $\hbar^0$ of Eq. \eqref{faNS}.\\

\section{Concluding remarks}

In this work we have  studied  the radial action for massive particles in three spherically symmetric spacetimes: the standard $d=4$ Schwarzschild background, its $d$-dimensional variation known as Schwarzschild-Tangherlini solution and the D3 branes spacetimes.

The radial action (and hence, the scattering angle, for example) can be integrated exactly in terms of special functions (elliptic functions or via a Lauricella representation in terms of generalized hypergeometric functions).
However, having a final result written in terms of elliptic functions is not very useful in the applications, because of numerous formal difficulties. 
For example, a large-$j$ expansion limit of the results needs expanding elliptic function both in  the modulus and the variable, a fact which complicates matters. Typically one expands the integrand and then obtains this sought for large-$j$ expansion limit. This expanded result is actually mostly used in the applications, but one has lost all the advantages of a resummed, exact result.
We have shown here that the resummation is still possible in the eikonal limit, starting from large-$j$ expanded result, generalizing to the massive particle case some property previously studied only in the massless case.

Furthermore, we have connected the radial action to the renormalized angular momentum $\nu$ (playing a role in the monodromy properties at infinity of the solutions of the radial equation for a massive scalar field), in the eikonal limit.

Also in this case,  previous results were limited to the massless case. 
It is worth to mention that to perform this computation we have used the qSW formalism, since the companion MST formalism has been only  developed for massless field perturbations.
The renormalized angular momentum in the qSW language corresponds to the $a$-cycle which we have analyzed here together with is dual $a_D$-cycle.

We expect that our findings will play a role for treating massive field perturbations of black holes.

We have examined with special attention the $d$-dimensional Schwarzschild-Tangherlini solution, proving resummation properties of the radial action (and the scattering angle as well) in terms of Fox-Wright functions. These results are expected to be of some help, especially in the dimensional regularization of otherwise divergent expressions of observables. 
Explicit applications, however, seem still challenging and left for future works.

\section*{Acknowledgments}

D.B. and G.D.R. acknowledge  membership to the Italian Gruppo Nazionale per
la Fisica Matematica (GNFM) of the Istituto Nazionale
di Alta Matematica (INDAM). 

\appendix

\section{Radial action in terms of elliptic integrals in the Schwarzschild spacetime}
\label{App_ellip}

The radial action, Eq. \eqref{regIr}, can be directly integrated in terms of elliptic integrals, namely
\begin{widetext}
\bea
\label{I_reg_ellip}
    I_r^{\rm reg}&=&C_0+C_{E,1} E\left(\left.\text{arcsin}(\kappa )\right|\gamma
   \right)+C_{F,1}F\left(\left.\text{arcsin}(\rho )\right|\mu \right)+C_{F,2}F\left(\left.\text{arcsin}(\nu )\right|\mu \right)+C_{F,3}F\left(\left.\text{arcsin}(\kappa )\right|\gamma
   \right)\nonumber\\
   &+&C_{\Pi,1}\Pi \left(\delta ;\left.\text{arcsin}(\kappa
   )\right|\gamma \right)+C_{\Pi,2}\Pi \left(\alpha ;\left.\text{arcsin}(\rho )\right|\mu
   \right)+C_{\Pi,3}\Pi \left(\alpha ;\left.\text{arcsin}(\sigma
   )\right|\mu \right)+C_{\Pi,4}\Pi \left(\beta ;\left.\text{arcsin}(\rho )\right|\mu
   \right)\nonumber\\
   &+&C_{\Pi,5}\Pi \left(\beta ;\left.\text{arcsin}(\nu )\right|\mu
   \right)\,,
\eea
\end{widetext}
where the exact expressions of the various coefficients are listed   in Table \ref{tab:table1} below, whereas their approximated expressions (in a large-$j$ expansion limit) are given by\\
\bea
C_0 
&=& \frac{1}{\sqrt{\gamma^2-1}j}+\frac{\gamma^2}{(\gamma^2-1)j^2}+O\left(\frac{1}{j^3}\right)\,,\nonumber\\
C_{E,1}
&= & -\frac{2i(\gamma^2-1)^{1/4}}{j^{1/2}}-\frac{2i\gamma^2}{(\gamma^2-1)^{1/4}j^{3/2}}+O\left(\frac{1}{j^{5/2}}\right)\,,\nonumber\\
C_{F,1}
   &=& -\frac{8i(2\gamma^2-1)}{j^2}-\frac{16(\gamma^2-1)^{1/4}(2\gamma^2-1)}{j^{5/2}}+O\left(j^{3}\right)\,,\nonumber\\
C_{F,2}
   &=& \frac{8i(2\gamma^2-1)}{j^2}+\frac{16(\gamma^2-1)^{1/4}(2\gamma^2-1)}{j^{5/2}}+O\left(\frac{1}{j^3}\right)\,,\nonumber\\
C_{F,3}
&=& \frac{i \sqrt{j}}{(\gamma^2-1)^{1/4}}+\frac{i(2\gamma^2-1)}{(\gamma^2-1)^{3/4}j^{1/2}}+O\left(\frac{1}{j^{3/2}}\right)\,,
\eea
and
\bea
C_{\Pi,1}
&=& -\frac{4i\gamma^2}{(\gamma^2-1)^{1/4}j^{3/2}}-\frac{4i\gamma^2(3\gamma^2-2)}{(\gamma^2-1)^{3/4}j^{5/2}}+O\left(\frac{1}{j^{7/2}}\right)\,,\nonumber\\
C_{\Pi,2}
   &=&-\frac{4\sqrt{2}(2\gamma^2-1)}{(\gamma^2-1)^{1/4}j^{3/2}}+\frac{8i\sqrt{2}(2\gamma^2-1)}{j^2}+O\left(\frac{1}{j^{5/2}}\right)\,,\nonumber\\
C_{\Pi,3}
   &=&\frac{4\sqrt{2}(2\gamma^2-1)}{(\gamma^2-1)^{1/4}j^{3/2}}-\frac{8i \sqrt{2}(2\gamma^2-1)}{j^2}+O\left(\frac{1}{j^{5/2}}\right)\,,\nonumber\\
C_{\Pi,4}
&=&\frac{4\sqrt{2}(2\gamma^2-1)}{(\gamma^2-1)^{1/4}j^{3/2}}-\frac{8i \sqrt{2}(2\gamma^2-1)}{j^2}+O\left(\frac{1}{j^{5/2}}\right)\,,\nonumber\\
C_{\Pi,5}
   &=&-\frac{4\sqrt{2}(2\gamma^2-1)}{(\gamma^2-1)^{1/4}j^{3/2}}+\frac{8i \sqrt{2}(2\gamma^2-1)}{j^2}+O\left(\frac{1}{j^{5/2}}\right)\,,\nonumber\\
\eea
having defined
\bea
\alpha&=&\frac{A_-}{A_+}\frac{B_+}{B_-}
\nonumber\\
&\sim& 1+\frac{2i(\sqrt{2}-2)(\gamma^2-1)^{1/4}}{\sqrt{j}}+O\left(\frac{1}{j}\right)\,,\nonumber\\
\beta&=& \frac{A_+}{A_-}\frac{B_+}{B_-}
\nonumber\\
&\sim& 1-\frac{2i(2+\sqrt{2})(\gamma^2-1)^{1/4}}{\sqrt{j}}+O\left(\frac{1}{j}\right)\,,\nonumber\\
\gamma&=&-\frac{2 \left(u_1-u_3\right)}{2 u_1+4 u_3-1}\nonumber\\
&\sim& -\frac{j}{4\sqrt{\gamma^2-1}}+\frac{1}{2}+O\left(\frac{1}{j}\right)\,,\nonumber\\
\delta&=&\frac{2 u_1-2 u_3}{1-2 u_3}\nonumber\\
&\sim & 1-\frac{4\gamma^2}{j^2}+O\left(\frac{1}{j^3}\right)\,,\nonumber
\eea
\bea
\mu&=&\left(\frac{B_+}{B_-}\right)^2
\nonumber\\
&\sim& 1-\frac{8i(\gamma^2-1)^{1/4}}{\sqrt{j}}+O\left(\frac{1}{j}\right)\,,\nonumber\\
\nu&=& \sqrt{\frac{A_+B_-}{A_-B_+}}
\nonumber\\
&\sim& 1-\frac{i(\sqrt{2}-2)(\gamma^2-1)^{1/4}}{\sqrt{j}}+O\left(\frac{1}{j}\right)\,,\nonumber\\
\sigma&=&\sqrt{\frac{A_+
   \left(\sqrt{u_3}-\sqrt{u_1+2u_3-\frac{1}{2}}\right)-u_1}{A_- B_+}}\nonumber\\
&\sim& 1-\frac{i(\sqrt{2}-2)(\gamma^2-1)^{1/4}}{\sqrt{j}}+O\left(\frac{1}{j}\right)\,,\nonumber\\
\rho&=&\sqrt{\frac{B_-}{B_+}}
\nonumber\\
&\sim& 1+\frac{2i(\gamma^2-1)^{1/4}}{\sqrt{j}}+O\left(\frac{1}{j}\right)\,,\nonumber\\
\kappa&=&i \sqrt{\frac{u_3}{u_1-u_3}}\nonumber\\
&\sim& \frac{i\sqrt{2}(\gamma^2-1)^{1/4}}{\sqrt{j}}+\frac{i(3\gamma^2-2)}{\sqrt{2}(\gamma^2-1)^{1/4}j^{3/2}}\nonumber\\
&+& O\left(\frac{1}{j^{5/2}}\right)\,,
\eea
where
\bea
A_\pm &=&i\sqrt{u_1-u_3}\pm \sqrt{u_3}\,, \nonumber\\
B_\pm &=&i\sqrt{u_1-u_3}\pm \sqrt{u_1+2u_3-\frac{1}{2}} \,,
\eea
and we used the property of the turning points $\sum_{i=1}^3u_i=\frac{1}{2}$ and we defined the elliptic integrals as
\bea
F(\phi|m)&=&\int_0^\phi(1-m\sin^2\theta)^{-1/2}d\theta\,,\nonumber\\
E(\phi|m)&=&\int_0^\phi(1-m\sin^2\theta)^{1/2}d\theta\,,\nonumber\\
\Pi(n;\phi|m)&=&\int_0^\phi(1-n\sin^2\theta)^{-1}(1-m\sin^2\theta)^{-1/2}\,.\nonumber\\
\eea

\begin{table*}  
\caption{\label{tab:table1}  List of the exact expressions for the coefficients entering Eq. \eqref{I_reg_ellip}.}
\begin{ruledtabular}
\begin{tabular}{ll}
$C_0$& $\frac{1}{2} \left(\frac{1-2 u_1}{2 u_1+2
   u_3-1}-\frac{u_3}{u_1}\right)$\\
$C_{E,1}$&$-\frac{i \sqrt{u_3} \sqrt{2 u_1+4
   u_3-1}}{\sqrt{u_1} \sqrt{2 u_1+2 u_3-1}}$\\
$C_{F,1}$&$-\frac{2 
   \left(\left(8 u_3-2\right) u_1^2+\left(8 u_3^2-6
   u_3+1\right) u_1-2 u_3^2+u_3\right)}{u_1^{3/2}
   \sqrt{2 u_1+2 u_3-1} \left(\sqrt{2} u_1-\sqrt{2}
   u_3+i \sqrt{u_1-u_3} \sqrt{2 u_1+4
   u_3-1}\right)}$\\
   &$\times  \sqrt{\frac{u_3 \left({-}6 u_3{+}2 \sqrt{2}
   \sqrt{{-}\left(\left(u_1{-}u_3\right) \left(2 u_1{+}4
   u_3{-}1\right)\right)}{+}1\right)}{4 u_1{+}2 u_3{-}1}}
   \sqrt{{-}\frac{\left(u_1{-}u_3\right) \left(2
   i\sqrt{u_1{-}u_3}+\sqrt{4 u_1{+}8 u_3{-}2}\right)}{2
   i\sqrt{u_1{-}u_3}{-}\sqrt{4 u_1{+}8 u_3{-}2}}}$\\
$C_{F,2}$ &$\frac{1}{{u_1^2 \left(2 u_1{+}2
   u_3{-}1\right) \left(2 u_1{-}2 u_3{+}i \sqrt{2}
   \sqrt{u_1{-}u_3} \sqrt{2 u_1{+}4 u_3{-}1}\right)}}\sqrt{{-}\frac{u_3 \left({-}2 u_1^2{+}u_1{+}u_3
   \left(2 u_3{-}1\right)\right)}{\left(u_1{-}2 u_3{+}2 i
   \sqrt{u_1{-}u_3} \sqrt{u_3}\right) \left(4 u_1{+}2
   u_3{-}1\right)}}$\\
   &$\times \Bigg[2  \left(\left(8 u_3{-}2\right)
   u_1^2{+}\left(8 u_3^2{-}6 u_3{+}1\right) u_1{-}2
   u_3^2{+}u_3\right)  \left(4
   u_3^{3/2}{-}4 i \sqrt{u_1{-}u_3} u_3-2 \sqrt{4 u_1{+}8
   u_3{-}2} u_3\right.$\\
   &$+ \left.2 \sqrt{{-}4 u_1{-}8 u_3{+}2}
   \sqrt{u_1{-}u_3} \sqrt{u_3}{+}u_1 \left(2 i
   \sqrt{u_1{-}u_3}{-}4 \sqrt{u_3}{+}\sqrt{4 u_1{+}8
   u_3{-}2}\right)\right)\Bigg]$\\
   &$ \times \sqrt{\frac{2 u_1-2
   u_3+\sqrt{u_3} \left(\sqrt{4 u_1+8 u_3-2}-2 i
   \sqrt{u_1-u_3}\right)+\sqrt{-4 u_1-8 u_3+2}
   \sqrt{u_1-u_3}}{\left(\sqrt{u_1-u_3}+i
   \sqrt{u_3}\right) \left(2 \sqrt{u_1-u_3}-i
   \sqrt{4 u_1+8 u_3-2}\right)}}$\\
$C_{F,3}$&$\frac{i \sqrt{u_3} \left(2 u_1+2
   u_3+1\right)}{\sqrt{u_1} \sqrt{2 u_1+2 u_3-1}
   \sqrt{2 u_1+4 u_3-1}}$\\
$C_{\Pi,1}$& $\frac{4 i \left(2 u_1-1\right) \sqrt{u_3}
   \left(u_1+u_3\right)}{\sqrt{u_1} \sqrt{2 u_1+2
   u_3-1} \sqrt{2 u_1+4 u_3-1}}$\\
$C_{\Pi,2}$&$\frac{2 \sqrt{u_1-u_3}  \left(\left(8 u_3-2\right)
   u_1^2+\left(8 u_3^2-6 u_3+1\right) u_1-2
   u_3^2+u_3\right)}{u_1^{3/2} \sqrt{2 u_1+2 u_3-1}
   \left(-i \sqrt{2} u_1+i \sqrt{2}
   u_3+\sqrt{u_1-u_3} \sqrt{2 u_1+4 u_3-1}\right)}$\\
   &$\times \sqrt{\frac{-6 u_3+2
   \sqrt{2} \sqrt{-\left(\left(u_1-u_3\right)
   \left(2 u_1+4 u_3-1\right)\right)}+1}{4 u_1+2
   u_3-1}} \sqrt{-\frac{\left(u_1-u_3\right)
   \left(2 \sqrt{u_1-u_3}-i \sqrt{4 u_1+8
   u_3-2}\right)}{2 \sqrt{u_1-u_3}+i \sqrt{4 u_1+8
   u_3-2}}}$\\
$C_{\Pi,3}$&$\frac{\sqrt{\frac{2 u_1^2{-}u_1{-}2
   u_3^2{+}u_3}{\left(u_1{-}2 u_3{+}2 i \sqrt{u_1{-}u_3}
   \sqrt{u_3}\right) \left(4 u_1{+}2 u_3{-}1\right)}}}{u_1^2 \left(2
   u_1+2 u_3-1\right) \left(-2 i u_1+2 i
   u_3+\sqrt{2} \sqrt{u_1-u_3} \sqrt{2 u_1+4
   u_3-1}\right)}\Bigg[2 
   \left(\left(8 u_3-2\right) u_1^2+\left(8 u_3^2-6
   u_3+1\right) u_1\right.$\\
   &$- \left.2 u_3^2+u_3\right)
    \left(-2 i
   u_1^2+\left(6 i u_3+\left(4 \sqrt{u_1-u_3}-2 i
   \sqrt{4 u_1+8 u_3-2}\right) \sqrt{u_3}+i
   \sqrt{-4 u_1-8 u_3+2} \sqrt{u_1-u_3}\right)
   u_1\right.$\\
   &$- \left.2 i u_3 \left(2 u_3+\left(-\sqrt{4 u_1+8
   u_3-2}-2 i \sqrt{u_1-u_3}\right)
   \sqrt{u_3}+\sqrt{-4 u_1-8 u_3+2}
   \sqrt{u_1-u_3}\right)\right)\Bigg]$\\
   &$\times \sqrt{\frac{2 u_1{-}2 u_3{+}\sqrt{u_3} \left(\sqrt{4
   u_1{+}8 u_3{-}2}{-}2 i \sqrt{u_1{-}u_3}\right){+}\sqrt{{-}4
   u_1{-}8 u_3{+}2}
   \sqrt{u_1{-}u_3}}{\left(\sqrt{u_1{-}u_3}{+}i
   \sqrt{u_3}\right) \left(2 \sqrt{u_1{-}u_3}{-}i
   \sqrt{4 u_1{+}8 u_3{-}2}\right)}}$\\
$C_{\Pi,4}$ & $-\frac{2 \sqrt{u_1-u_3}  \left(\left(8 u_3-2\right)
   u_1^2+\left(8 u_3^2-6 u_3+1\right) u_1-2
   u_3^2+u_3\right)}{u_1^{3/2} \sqrt{2 u_1+2 u_3-1}
   \left(-i \sqrt{2} u_1+i \sqrt{2}
   u_3+\sqrt{u_1-u_3} \sqrt{2 u_1+4 u_3-1}\right)}$\\
   &$ \times \sqrt{\frac{-6 u_3+2
   \sqrt{2} \sqrt{-\left(\left(u_1-u_3\right)
   \left(2 u_1+4 u_3-1\right)\right)}+1}{4 u_1+2
   u_3-1}} \sqrt{-\frac{\left(u_1-u_3\right)
   \left(2 \sqrt{u_1-u_3}-i \sqrt{4 u_1+8
   u_3-2}\right)}{2 \sqrt{u_1-u_3}+i \sqrt{4 u_1+8
   u_3-2}}}$\\
$C_{\Pi,5}$&$-\frac{1}{u_1^2 \left(2 u_1+2
   u_3-1\right) \left(2 u_1-2 u_3+i \sqrt{2}
   \sqrt{u_1-u_3} \sqrt{2 u_1+4 u_3-1}\right)}\Bigg[2  \left(\left(8 u_3-2\right)
   u_1^2+\left(8 u_3^2-6 u_3+1\right) u_1\right.$\\
   &$- \left.2
   u_3^2+u_3\right)  \left(2
   u_1^2{+}\left({-}6 u_3{+}2 \left(\sqrt{4 u_1{+}8
   u_3{-}2}{+}2 i \sqrt{u_1{-}u_3}\right) \sqrt{u_3}{-}i
   \sqrt{2} \sqrt{u_1{-}u_3} \sqrt{2 u_1{+}4
   u_3{-}1}\right) u_1\right.$\\
   &$+ \left.2 \left(\sqrt{u_3}{-}i
   \sqrt{u_1{-}u_3}\right) u_3 \left(2
   \sqrt{u_3}{-}\sqrt{4 u_1{+}8
   u_3{-}2}\right)\right)\Bigg]\sqrt{\frac{\left(u_1-u_3\right) \left(2
   u_1+2 u_3-1\right)}{\left(u_1-2 u_3+2 i
   \sqrt{u_1-u_3} \sqrt{u_3}\right) \left(4 u_1+2
   u_3-1\right)}}$\\
   &$\times \sqrt{\frac{-2
   u_1+\left(\sqrt{u_3}+i \sqrt{u_1-u_3}\right)
   \left(2 \sqrt{u_3}-\sqrt{4 u_1+8
   u_3-2}\right)}{\left(-\sqrt{u_3}+i
   \sqrt{u_1-u_3}\right) \left(\sqrt{4 u_1+8
   u_3-2}+2 i \sqrt{u_1-u_3}\right)}}$\\
\end{tabular}
\end{ruledtabular}
\end{table*}

\end{document}